\documentclass[preprint2]{aastex63}

\usepackage{graphicx}
\usepackage{natbib}
\usepackage{booktabs}
\usepackage{amsmath, physics}

\newcommand\TB{T_B}

\shorttitle{2---4~GHz Continuum Emission from $\epsilon$ Eridani}
\shortauthors{Suresh et al.}

\begin{document}

\title{Detection of 2---4~GHz Continuum Emission from $\epsilon$ Eridani}

\correspondingauthor{Akshay Suresh}
\email{as3655@cornell.edu}

\author[0000-0002-5389-7806]{A.~Suresh}
\affiliation{Cornell Center for Astrophysics and Planetary Science, 
             and Department of Astronomy,
             Cornell University, Ithaca, NY 14853, USA}         
             
\author[0000-0002-2878-1502]{S.~Chatterjee}
\affiliation{Cornell Center for Astrophysics and Planetary Science, 
             and Department of Astronomy,
             Cornell University, Ithaca, NY 14853, USA}
             
\author[0000-0002-4049-1882]{J.~M.~Cordes}
\affiliation{Cornell Center for Astrophysics and Planetary Science, 
             and Department of Astronomy,
             Cornell University, Ithaca, NY 14853, USA}
             
\author[0000-0002-0713-0604]{T.~S.~Bastian}
\affiliation{National Radio Astronomy Observatory, 520 Edgemont Road, Charlottesville, VA 22903, USA}             

\author[0000-0002-7083-4049]{G.~Hallinan}
\affiliation{Cahill Center for Astronomy, California Institute of Technology, Pasadena, CA 91125, USA}     
   


\begin{abstract}
The nearby star $\epsilon$~Eridani has been a frequent target of radio surveys for stellar emission and extraterrestial intelligence. Using deep 2---4~GHz observations with the Very Large Array, we have uncovered a $29 \ \mu {\rm Jy}$ compact, steady continuum radio source coincident with $\rm \epsilon \ Eridani$ to within $0\farcs06$ ($\lesssim 2\sigma$; 0.2 au at the distance of the star). Combining our data with previous high frequency continuum detections of $\rm \epsilon \ Eridani$, our observations reveal a spectral turnover at $\rm 6 \ GHz$. We ascribe the 2---6~GHz emission to optically thick, thermal gyroresonance radiation from the stellar corona, with thermal free-free opacity likely becoming relevant at frequencies below 1~GHz. The steep spectral index ($\alpha \simeq 2$) of the 2---6~GHz spectrum strongly disfavors its interpretation as stellar wind-associated thermal bremsstrahlung ($\alpha \simeq 0.6$). Attributing the entire observed 2---4~GHz flux density to thermal free-free wind emission, we thus, derive a stringent upper limit of $3 \times 10^{-11} \ M_{\odot} \ {\rm yr}^{-1}$ on the mass loss rate from $\rm \epsilon \ Eridani$. Finally, we report the non-detection of flares in our data above a $5\sigma$ threshold of $\rm 95 \ \mu Jy$. Together with the optical non-detection of the most recent stellar maximum expected in 2019, our observations postulate a likely evolution of the internal dynamo of $\rm \epsilon \ Eridani$.
\end{abstract}

\keywords{Stellar coronae (305), K stars (878), Radio continuum emission (1340), Stellar atmospheres (1584)}

\section{Introduction}\label{sec1}
Radio surveys of stellar systems have garnered renewed interest in the last two decades, motivated partly by searches for exoplanetary radio emission. Among the targets in early radio searches for extraterrestrial intelligence \citep{Drake1961, Henry1995, Tarter1996} and microwave coronal emission \citep{Gary1981} is $\rm \epsilon \ Eridani $ (hereafter, $\rm \epsilon \ Eri $), a K2V dwarf at a distance of 3.2 pc \citep{GaiaDR2}. With a mass loss rate at least 30 times \citep{Wood2002} that of the Sun, $\rm \epsilon \ Eri$ displays more vigorous chromospheric and coronal activity levels than the present Sun. Properties of $\rm \epsilon \ Eri $ are summarized in Table \ref{tab1}. \\

\begin{deluxetable}{lCc}
\tablecaption{Properties of $\rm \epsilon \ Eri$.\label{tab1}}
\tablewidth{0pt}
\tablehead{
\colhead{Property} & \colhead{Value} & \colhead{Refs.}
}
\startdata
Spectral type & \rm K2V & (1) \\
Age (Myr) & 200-800 & (2)  \\
Distance (pc) & 3.20 \pm 0.03 & (3) \\
Mass ($M_{\odot}$) & 0.83 \pm  0.01 & (4)  \\
Radius ($R_{\odot}$) & 0.74 \pm 0.01 & (4) \\
Effective temperature (K) & 5100 \pm 16  \phn \phn & (4) \\
Surface magnetic field (G)  & 186 \pm 47  \phn & (5) \\
X-ray luminosity\tablenotemark{$\dagger$} ($\rm erg \ s^{-1}$) & 10^{28.5} & (6) \\
Radio luminosity\tablenotemark{$\ddagger$} ($\rm erg \ s^{-1}$) & 10^{12} & (7)
\enddata
\tablenotetext{\dagger}{ X-ray luminosity in the $0.2 - 20$ keV band.}
\tablenotetext{\ddagger}{ Radio luminosity in the $4-8$ GHz band.}
\tablerefs{ (1)~\citet{Keenan1989}, (2)~\citet{Mamajek2008}, (3)~\citet{GaiaDR2}, (4)~\citet{Bonfanti2015}, (5)~\citet{Lehmann2015}, (6)~\citet{Johnson1981}, (7)~\citet{Bastian2018}.}
\vspace{-8mm}
\end{deluxetable}

$\rm \epsilon \ Eri $ hosts an adolescent (200---800~Myr, \citealt{Mamajek2008}) planetary system that has aroused great scientific interest over the past decades. The confirmed members of this system include a Jovian mass exoplanet, $\rm \epsilon \ Eri \ b$, with a semi-major axis, $a \simeq 3.4$~au \citep{Hatzes2000, Mawet2019}, and an outer debris disk \citep{Greaves2014, ChavezDagostino2016} at $a \simeq 64$~au. Additional planets \citep{Quillen2002}, warm inner disks at $a \sim 3$~au and $ \sim 20$~au \citep{Backman2009}, and in situ dust-producing belts \citep{Su2017} at $a \leq 25$~au have also been proposed, but are not conclusively established. \\

Table \ref{tab2} summarizes detections of continuum emission from $\rm \epsilon \ Eri$ over 2---300~GHz \citep{Lestrade2015, MacGregor2015, ChavezDagostino2016,  Booth2017, Bastian2018, Rodriguez2019}. While the millimeter wave emission is attributed to optically thick chromospheric thermal emission \citep{MacGregor2015, Booth2017}, the remarkably flat 6---50~GHz spectrum is believed to be produced either via optically thin, thermal free-free emission from a stellar wind \citep{Rodriguez2019}, or through a combination of optically thick, thermal free-free and thermal gyroresonance radiation  \citep{Bastian2018} from the stellar corona. Aside from these continuum emissions, \citet{Bastian2018} also serendipitously detected a five-minute-long, 1~mJy flare at 2---4~GHz. While the identification of $\rm \epsilon \ Eri $ as a moderately active star favors a stellar origin for their observed flare, its $\sim 1''$ separation from $\rm \epsilon \ Eri$ raises the possibility of a non-stellar provenance.\\

\begin{deluxetable*}{l c c c c c}
\tablecolumns{4}
\tablewidth{0pc}
\tablecaption{Summary of continuum emission detected from $\rm \epsilon \ Eri$ at different frequency bands. \label{tab2}}
\tablehead{                        
Instrument &  Epoch  & Band & Flux Density  & $ \chi_{c}$\tablenotemark{$\dagger$} & Reference \\                
  &  &  (GHz) & ($\mu$Jy)  & ($\%$)  & }
\startdata
VLA    & 2019 Aug 03 &  $2 - 4$ & 30.3 $\pm$ 3.1 &  $<$ 25 & This paper \\
VLA    & 2019 Sep 05 &  $2 - 4$ & 28.4 $\pm$ 4.9 & $< $ 25 & This paper \\
VLA    & 2016 Mar 01 &  $2 - 4$ & $<$65\tablenotemark{$\ddagger$}  & \nodata & \citet{Bastian2018} \\
VLA    & 2016 Jan 21 &  $4 - 8$ &  \phn \phn 83 $\pm$ 16.6  & $<$ 50 & \citet{Bastian2018} \\
VLA    & 2013 May 18 &  \phn $8 - 12$ & 66.8 $\pm$ 3.7  & $<$ 14 & \citet{Bastian2018} \\
VLA    & 2013 May 19 &  \phn $8 - 12$ & 70.3 $\pm$ 2.7  & $<$ 10 & \citet{Bastian2018} \\
VLA    & 2013 Apr 20 & $12 - 18$ & 81.2 $\pm$ 6.6  & $<$ 20 & \citet{Bastian2018} \\
VLA    & 2017 Jun 15 & $29 - 37$ & \phn 70 $\pm$ 11  & $<$ 22 & \citet{Rodriguez2019} \\
ATCA    & 2014 Jun 25 to Aug 05 & $42 - 46$ & 66.1 $\pm$ 8.7  & \nodata & \citet{MacGregor2015} \\
SMA    & 2014 Jul 28 to Nov 19 &   $217 - 233$ & 1060 $\pm$ 300  & \nodata & \citet{MacGregor2015} \\
ALMA   & 2015 Jan 17 to Jan 18 & $226 - 234$  & 820 $\pm$ 68  & \nodata & \citet{Booth2017}  \\
IRAM   & 2007 Nov 16 to Dec 04 & $210 - 290$ &   1200 $\pm$ 300  & \nodata & \citet{Lestrade2015} \\
LMT   & 2014 Nov 01 to Dec 31 & $245 - 295$  &   2300 $\pm$ 300  & \nodata & \citet{ChavezDagostino2016}
\enddata
\tablenotetext{\dagger}{ 2.5$\sigma$ upper limit on circular polarization fraction, $\chi_c =  \rm |V|/I $.}
\tablenotetext{\ddagger}{ 2.5$\sigma$ upper limit on continuum flux density.}
\end{deluxetable*}

With the intent of localizing analogous flares and conclusively associating these with a stellar, planetary, or alternate origin, we performed high spatial resolution interferometric observations of  $\rm \epsilon \ Eri$. Our observations mark the first detection of 2---4~GHz quiescent continuum emission from $\rm \epsilon \ Eri$, although we find no flares in our data. Section~\ref{sec2} describes our observing setup and data analysis methods. We present our results in Section~\ref{sec3}, and discuss their physical implications in Section~\ref{sec4}. Finally, in Section~\ref{sec5}, we summarize our findings and present the conclusions from our study.

\section{Observations and Data Processing}\label{sec2}
Observations of $\epsilon$ Eri  in the 2---4~GHz frequency band were acquired using the Karl G. Jansky Very Large Array (VLA) on 2019~August~3 (epoch~1) and 2019~September~5 (epoch~2). During these epochs, the VLA was in its most extended configuration, yielding an angular resolution of $\simeq 0\farcs5$ at 2---4~GHz. Each observation spanned six hours, of which nearly five hours were spent on $\rm \epsilon \ Eri$. 3C138 was the flux density and bandpass calibrator, while the VLBA source J0331$-$1051 \citep{Beasley2002} was the complex gain calibrator in our observations. To enable precise astrometry, every ten-minute scan (phase center: $\alpha (\rm FK5 \ J2000) =$ $03^{\rm h}32^{\rm m}55\fs84, \ \delta (\rm FK5 \ J2000) = -09\degr 27\arcmin 29\farcs 73$) on $\epsilon$ Eri was interleaved between one-minute scans on J0331$-$1051. Since our calibration cycles are shorter than the typical VLA cycle time of 15~minutes at 2---4~GHz, we expect our VLA observations to yield more accurate gain phase solutions in comparison to a standard VLA observing program.\\ 

To verify our complex gain solutions, we additionally performed three two-minute scans on the VLA calibrator J0339$-$0146 at each epoch. However, the gain solutions derived from J0331$-$1051 result in a systematic error of $0\farcs05 - 0\farcs1$ on the position of J0339$-$0146 ($\sim 5\degr - 10\degr$ away from J0331$-$1051). Considering that J0331$-$1051 is separated from $\epsilon$~Eri by $\lesssim 1\degr$ on the sky, we use the complex gain solutions derived from J0331$-$1051 to calibrate our $\epsilon$~Eri data.\\

The primary motivation for our study was the detection of circularly polarized flares from the $\epsilon$~Eri system. The VLA uses native circularly polarized feeds at 2---4~GHz, thereby, circumventing the need for observing a polarization calibrator. \\
\begin{figure*}[ht!]
\includegraphics[width=.48\textwidth]{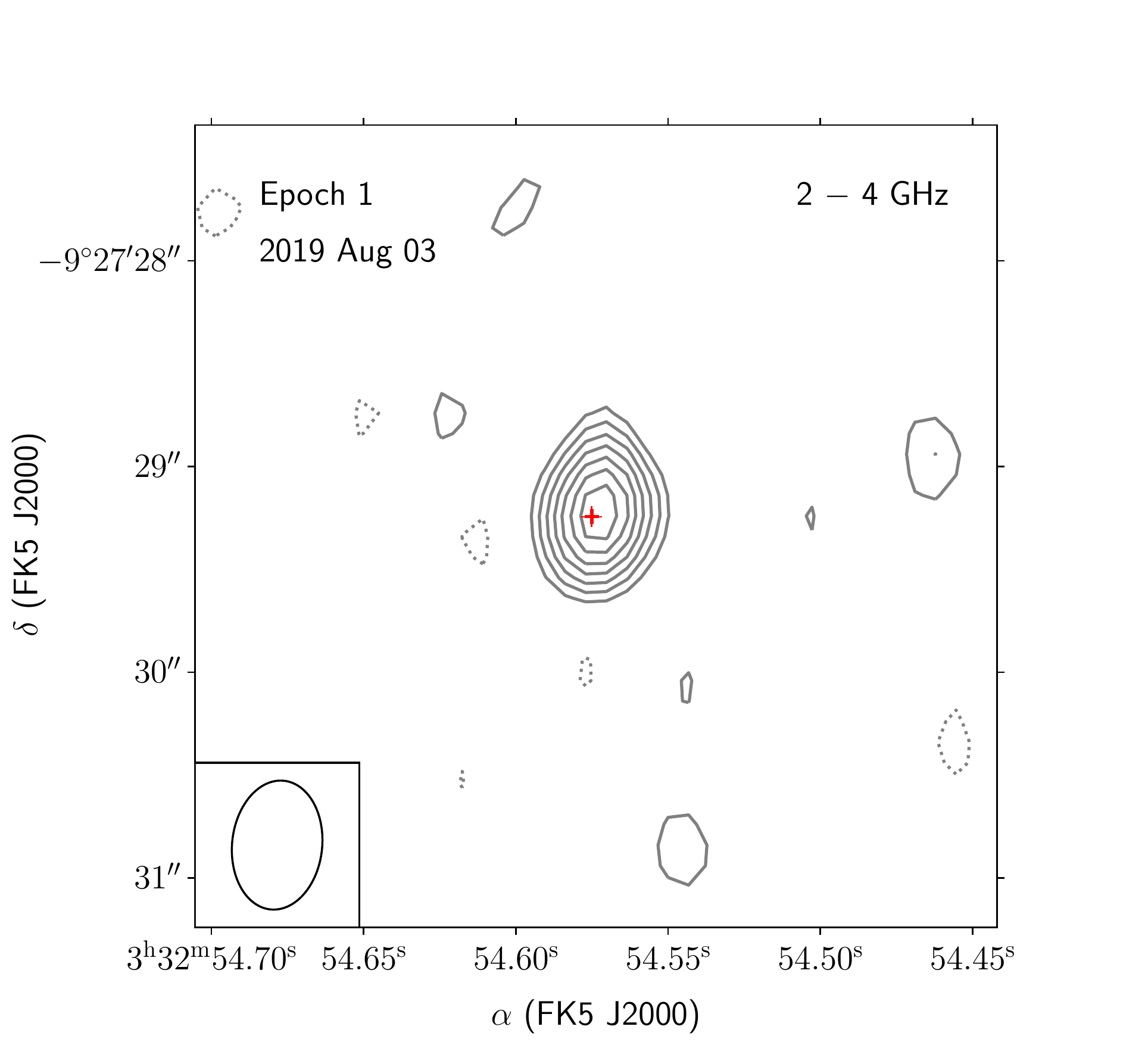}\quad
\includegraphics[width=.48\textwidth]{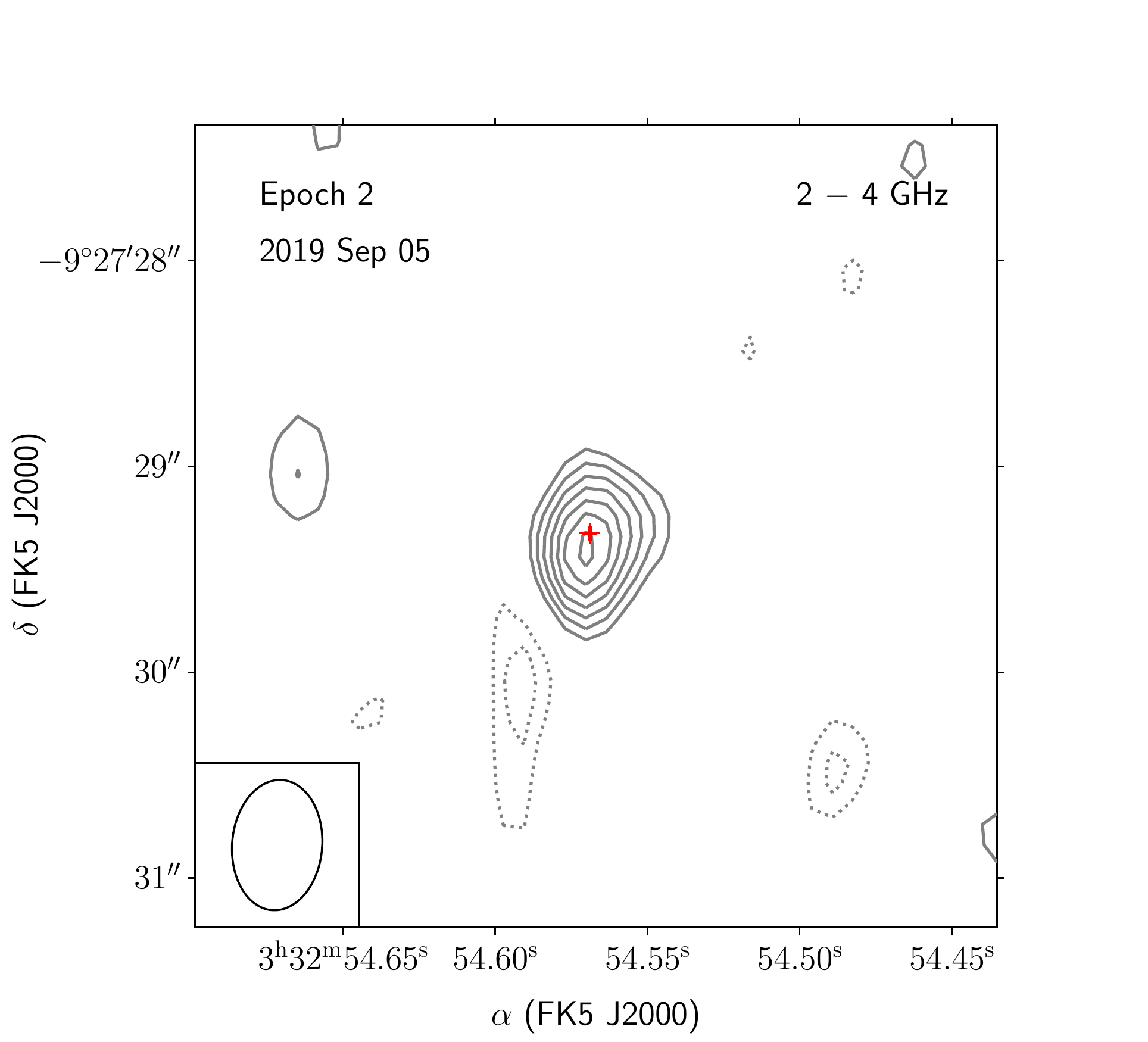}
\caption{Detection of 2---4~GHz continuum emission associated with $\rm \epsilon \ Eri$ at epochs 1 (left) and 2 (right). From outer to inner, the solid gray contours label 2, 3, 4, 5, 6, 7, and 8 times the rms intensity, $\sigma_{I_{\nu}} = 3.4  \ \mu$Jy~beam$^{-1}$. The dotted contours mark $-3\sigma_{I_{\nu}}$ and $-2\sigma_{I_{\nu}}$ levels. The red crosshairs in each panel indicate the {\it Gaia} DR2 position (with $1\sigma$ errors) of $\rm \epsilon \ Eri$ corrected for proper motion and annual parallax to the respective epochs. The black solid ellipse at the bottom left corner of each panel illustrates the VLA synthesized beam of size $0\farcs 63 \times  0\farcs 44$. \label{fig1}}
\end{figure*}

The data products analyzed in this work were processed using the standard VLA Stokes--I continuum calibration pipeline supplied with the Common Astronomy Software Applications (CASA, \citealt{CASA}) package. No additional post-pipeline processing or self-calibration was performed. Clipping bandpass edges and excising known satellite-associated radio frequency interference (RFI) in the 2.3---2.4~GHz band, approximately 90\% of our 2---4~GHz bandwidth is usable across 27 and 24 ``good'' antennas during epochs~1~and~2 respectively. We mapped the data from each epoch out to the first null of the primary beam, and built a model comprised of background sources in the field-of-view. Utilizing the model-subtracted visibilities \citep{Chiti2016}, series of radio images centered on $\epsilon \text{ Eri}$ were then constructed over different time scales ranging from a minute (duration of \citet{Bastian2018} flare) to five hours (net integration time on $\epsilon \text{ Eri}$ per epoch). \\

Briggs weighting \citep{Briggs1995} encapsulates a smooth transition from uniform weighting (finest angular resolution, lowest sensitivity) to natural weighting (greatest sensitivity, coarsest angular resolution) of complex visibilities using a robust parameter than can be varied continuously between $-2$ (uniform weighting) and $+2$ (natural weighting). For achieving the best compromise between sensitivity and angular resolution in our images, we adopted Briggs weighting  with a robust parameter of zero. 

\begin{deluxetable*}{cCcccc}
\tablecaption{Results of a 2D elliptical Gaussian fit to the 2---4~GHz radio counterpart of $\epsilon$~Eri. \label{tab3}}
\tablewidth{0pt}
\tablehead{
\colhead{ } & \colhead{ } & \multicolumn{2}{c}{Positions (FK5 J2000)\tablenotemark{$\dagger$}} & \colhead{ } & \colhead{ } \\
 \cmidrule(lr){3-4}
Epoch & \rm Start \ MJD & $\alpha_{\rm centroid}$ & $\delta_{\rm centroid}$ & Flux density\tablenotemark{a} & S/N\tablenotemark{b} \\
(number) &  & (h m s)  & ($\degr \ \arcmin \ ''$)  & ($\mu$Jy) & 
}
\startdata
1 & 58698.45 & 03 32 54.573(1) & $-$09 27 29.22(2) & $30.3 \pm 3.1$ & $9.02$\\
2 & 58731.37 & 03 32 54.568(2)  & $-$09 27 29.38(3) & $28.4 \pm 4.9$ & $8.45$ 
\enddata
\tablenotetext{\dagger}{Parenthesized numbers reflect uncertainties in the last significant digits.}
\tablenotetext{a}{Flux density uncertainties include not only internal fitting errors, but also a $5\%$ flux calibration error arising from the flux density uncertainty of our amplitude calibrator 3C138.}
\tablenotetext{b}{Detection signal-to-noise (S/N) is computed as the ratio of the peak spectral intensity ($I_{\nu}^{\rm peak}$) of the 2D Gaussian fit to the rms image intensity ($ \sigma_{ I_{\nu}}$).}
\end{deluxetable*}
\begin{figure}[th!]
\centering
\includegraphics[width=.48\textwidth, trim={0.25cm 0.1cm 1.4cm 1cm}, clip]{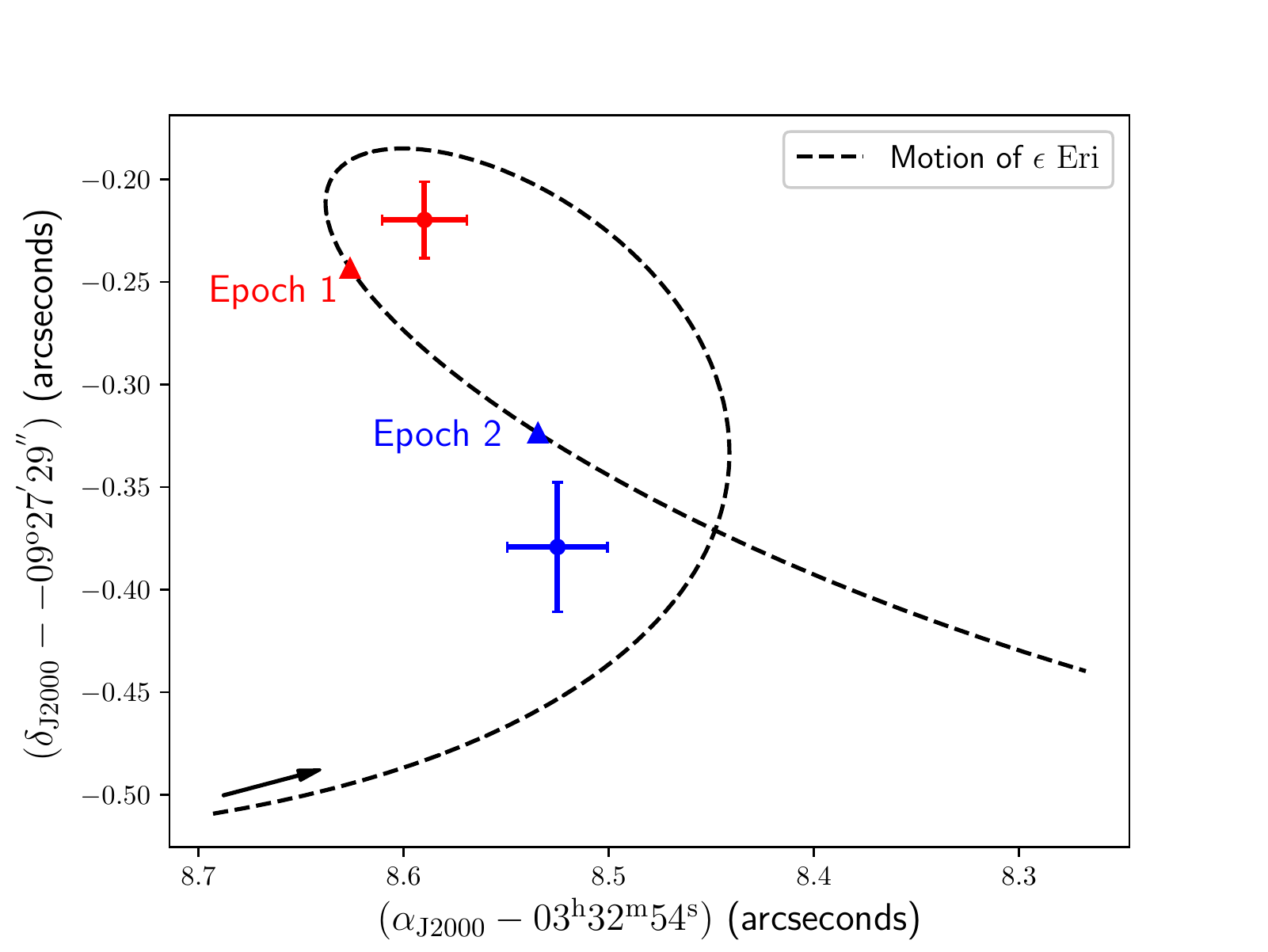}
\caption{Comparison of radio centroid locations (crosshairs showing $1\sigma$ errors) with the {\it Gaia} DR2 positions (triangular markers) of $\rm \epsilon \ Eri$ at epochs 1 (red) and 2 (blue). Positions are shown as offsets relative to the fiducial coordinates, $\alpha {\rm (J2000)} = 03^{\rm h}32^{\rm m}54^{\rm s}$ and $\delta {\rm (J2000)} = -09\degr27'29''$. The black dashed curve labels the path of $\rm \epsilon \ Eri$ on the sky over a ten-month time span, accounting for proper motion and annual parallax. The black solid arrow indicates the direction of motion of $\rm \epsilon \ Eri$ along its path. The {\it Gaia} DR2 positions of $\rm \epsilon \ Eri$ at our observing epochs have 3~mas position uncertainties along $\alpha$ and $\delta$ that are not visible on the plotted axes scales. \label{fig2}}
\end{figure}

\section{Results}\label{sec3}
Continuum images of the central $4'' \times 4''$ neighborhood of $\epsilon \text{ Eri}$ were generated by integrating over the entire 2---4~GHz band (after flagging bad channels and clipping bandpass edges) and the full exposure time on $\epsilon \text{ Eri}$ at each epoch. The resulting root-mean-squared (rms) noise estimated from a source-free region in these $4'' \times 4''$ cleaned images is $ \sigma_{I_{\nu}} \simeq 3.4$~$\mu$Jy~beam$^{-1}$. As shown in Figure \ref{fig1}, an unresolved source of flux density, $S_{\nu} \simeq 29 \rm \ \mu Jy$, is evident at both observing epochs. We use the CASA task {\tt imfit} to characterize the properties of this source via a 2D elliptical Gaussian fit. Table~\ref{tab3} summarizes the fit results, wherein we additionally incorporate the $5\%$ flux density uncertainty of 3C138 \citep{Perley2017} in our quoted radio source flux density errors. We introduce the systematic errors, $\rm \epsilon_{\alpha} = 19.2 \ mas$, and $\rm \epsilon_{\delta} = -10.4 \ mas$, to account for both the position uncertainty of the phase calibrator J0331$-$1051 in our VLA images, and the relative tie of the J0331$-$1051 VLBA frame \citep{Beasley2002} to our VLA coordinate frame. The net uncertainties on the absolute position of the radio centroid, $(\alpha_{\rm centroid}, \ \delta_{\rm centroid})$, are computed as the quadrature sum of the internal fitting errors and the systematic errors ($\epsilon_{\alpha}, \epsilon_{\delta}$). \\

\begin{figure*}[ht!]
\includegraphics[width=.48\textwidth]{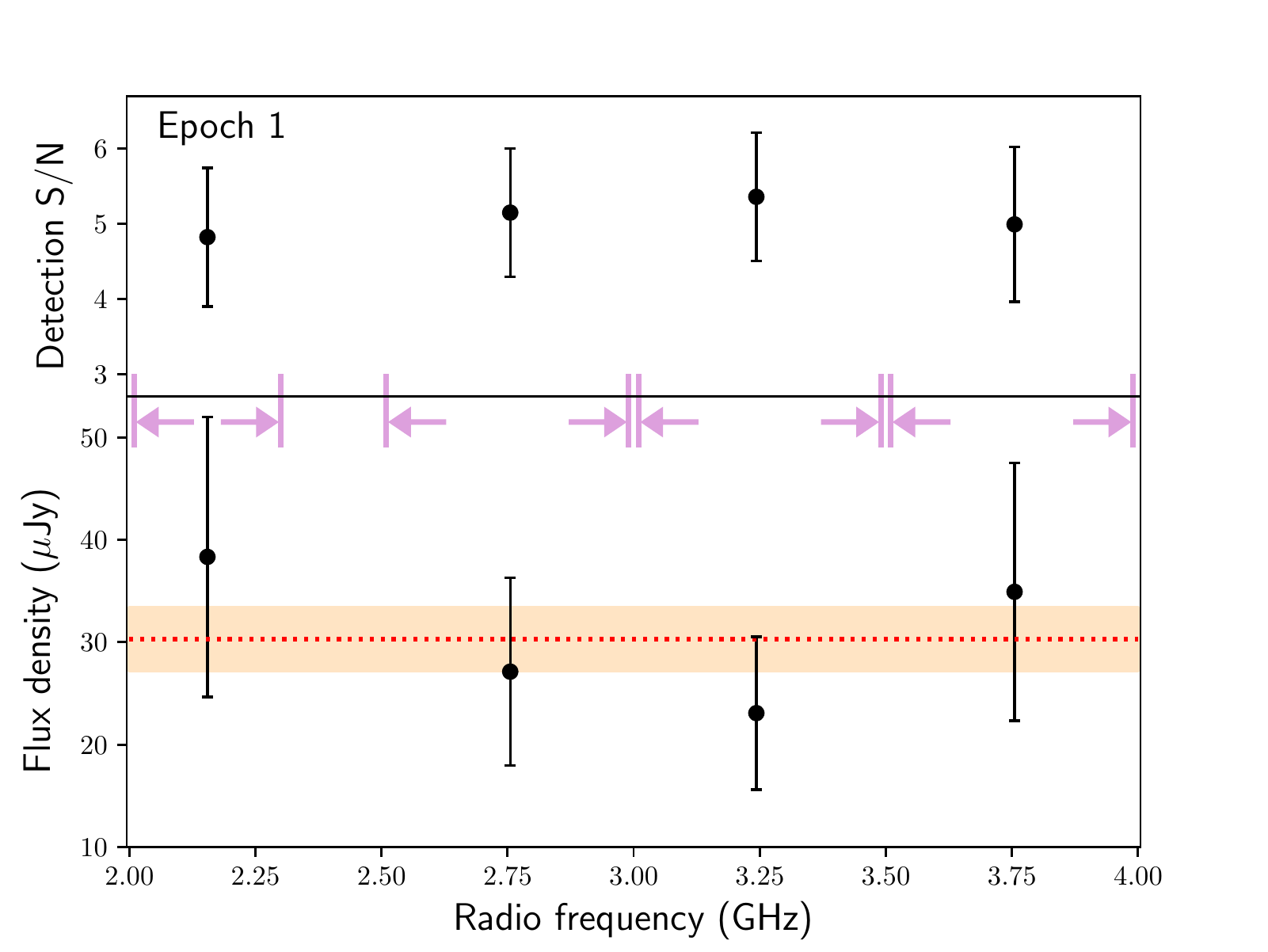}\quad
\includegraphics[width=.48\textwidth]{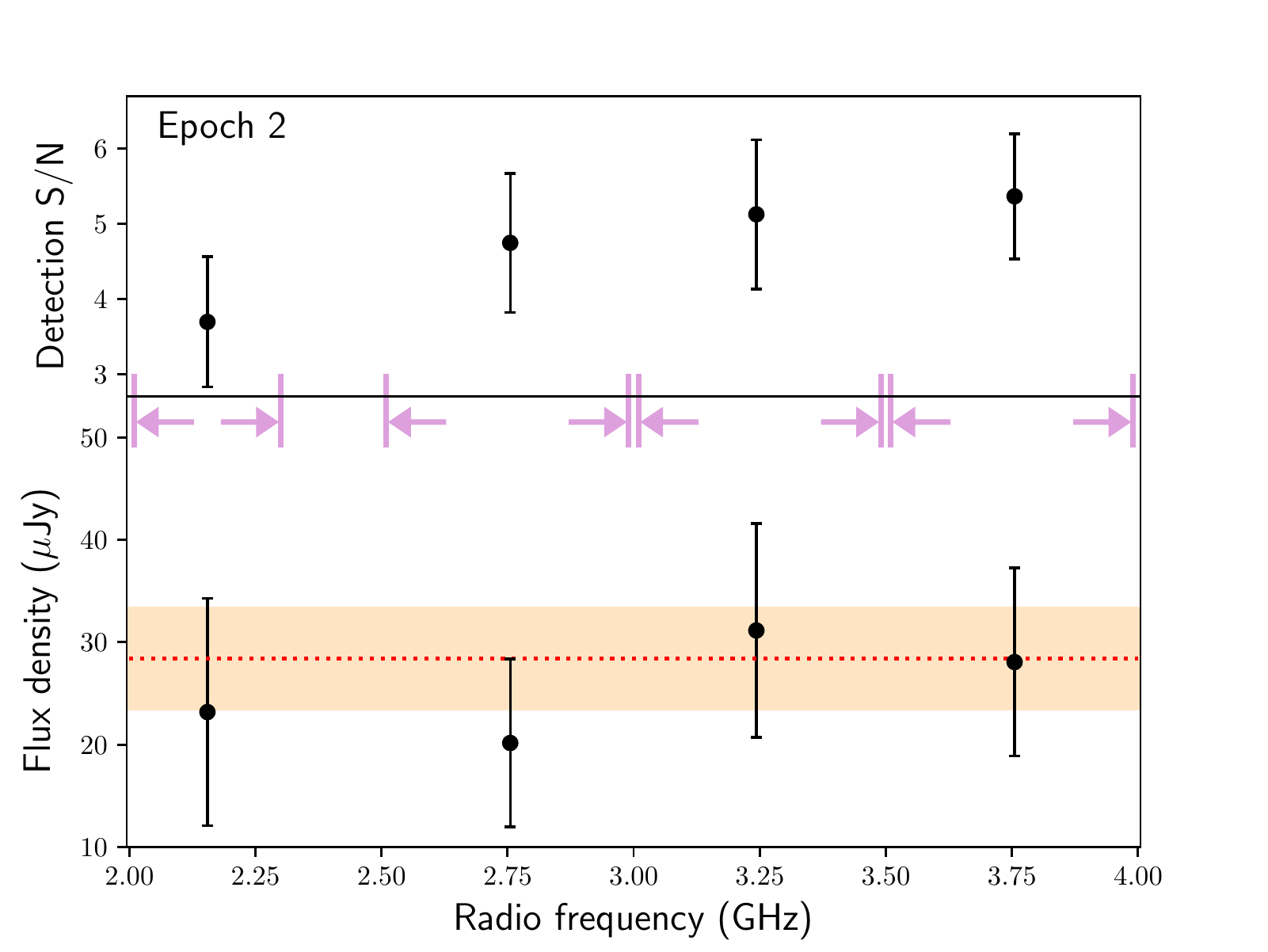}
\caption{Emission spectrum (lower panels) of the radio counterpart of $\epsilon$~Eri during epochs 1 (left) and 2 (right). The light pink vertical lines and horizontal arrows in each panel highlight subband edges. Note that the RFI-affected 2.3---2.4~GHz band is excluded from our analysis. The red dotted horizontal lines in bottom panels mark the frequency-averaged source flux density reported in Table \ref{tab3}. The orange band covers flux density values within $\pm 1\sigma$ of the red dotted line. Detection S/N values (top panels) at subband center frequencies are calculated as the ratio of the peak source intensity ($I_{\nu}^{\rm peak}$) to the rms image intensity ($\sigma_{I_{\nu}}$).
All plotted errorbars indicate $1\sigma$ errors on their respective quantities.\label{fig3}}
\end{figure*}

Figure \ref{fig2} compares $(\alpha_{\rm centroid}, \ \delta_{\rm centroid})$ at our two epochs with the  {\it Gaia} DR2 \citep{GaiaMission, GaiaDR2} positions of $\epsilon \text{ Eri}$ corrected for proper motion and annual parallax to the respective epochs. For a given epoch, let $\alpha_{\rm offset}$, $\delta_{\rm offset}$, and $\xi$ denote, respectively, the right ascension offset, the declination offset, and the net angular separation of our radio centroid from the {\it Gaia} DR2 position of $\epsilon \text{ Eri}$. We use standard error propagation rules for estimating uncertainties on $\alpha_{\rm offset}$, $ \delta_{\rm offset}$, and $\xi$. \\

For epoch 1, we have:
\begin{align}
\alpha_{\rm offset} &= \rm (-36.3 \pm 21.0 ) \ mas, \label{eqn1} \\
\delta_{\rm offset} &= \rm (23.6 \pm 18.9 ) \ mas, \label{eqn2} \\
\xi &= \rm (43.3 \pm 28.2) \ mas. \label{eqn3}
\end{align}
The values of the corresponding quantities for epoch 2 are:
\begin{align}
\alpha_{\rm offset} &= \rm (-9.5 \pm 24.7 ) \ mas, \label{eqn4} \\
\delta_{\rm offset} &= \rm (-55.5 \pm 31.6 ) \ mas, \label{eqn5}\\
\xi &= \rm (56.3 \pm 40.1) \ mas. \label{eqn6}
\end{align}
All of the above offsets are consistent with being zero at the $1-2\sigma$ level. We also consider the possibility that our detected emission is associated with the confirmed planet $\rm \epsilon \ Eri \ b$. According to \citet{Benedict2006}, $\rm \epsilon \ Eri \ b$ executes a $\sim$ 7-year orbit with semi-major axis, $a \simeq 3.4$~AU, eccentricity, $e \simeq 0.7$, and orbital inclination, $i = 30\fdg1$. Therefore, at the $\rm 3.2 \ pc$ distance \citep{GaiaDR2} of $\rm \epsilon \ Eri$, $\rm \epsilon \ Eri \ b$ must lie at least $0\farcs29$ away from its host star. On the contrary, \citet{Mawet2019} favor an edge-on orbital orientation for $\rm \epsilon \ Eri \ b$. However, the rapid orbital phase coverage of the emission centroid between our epochs implies an orbital period of 64---114~days (assuming circular orbit in the plane of the sky), which cannot be attributed to $\rm \epsilon \ Eri \ b$. In the absence of evidence for an alternate close-in body within this system, we conclude that our observed emission is coincident with $\rm \epsilon \ Eri$.

\subsection{Source Emission Properties}\label{sec3.1}
We find no evidence of Stokes$-$V emission associated with our 2---4~GHz continuum source, thereby constraining its circular polarization fraction, $\chi_c = \rm |V|/I \lesssim 25\%$, at the $2.5\sigma$ level. We characterize the source spectrum by performing epoch-integrated imaging of the $\rm \epsilon \ Eri$ field in four contiguous subbands covering 2---4~GHz. At each subband, we estimate the source flux density using the CASA task {\tt imfit}, but with the 2D Gaussian peak constrained to the coordinates $(\alpha_{\rm centroid}$, $\delta_{\rm centroid})$ inferred from the band-averaged image. As illustrated in Figure \ref{fig3}, the spectrum is consistent with being flat over 2---4~GHz at both observing epochs. \\

Implementing a similar procedure along the time axis, we find that the  radio source displays no significant intra-epoch variability. Imaging our observations separately in ten-minute and one-minute intervals, we find no evidence of emission above $5\sigma_{I_{\nu}}$ in our radio images, thus, placing an upper limit, $S_{\nu} \leq 95 \ \mu \rm Jy$, on any undetected weak flares in our data. \\ \\

\section{Discussions}\label{sec4}
Our VLA observations have yielded the first detection of quiescent continuum emission from $\rm \epsilon \ Eri$ in the 2---4~GHz band. Assuming a source size, $r_s \approx R_* = 0.74 \ R_{\odot}$ \citep{Bonfanti2015}, at the $d = 3.2 \ {\rm pc}$ distance \citep{GaiaDR2} of $\rm \epsilon \ Eri$, the observed $S_{\nu} \simeq 29 \ \mu \rm Jy$ implies a brightness temperature,
\begin{align}\label{eqn7}
\TB \simeq 1.2 \ {\rm MK} \ \left( \frac{S_{\nu, \ \rm 29 \ \mu Jy} \ d_{3.2 \ {\rm pc}}^2}{\nu_{3 \ {\rm GHz}}^2 \  r_{R_*}^2} \right). 
\end{align}
Allowing for a flat spectrum over $\rm 2-4 \ GHz$, $\TB \simeq 0.7 - 2.7 \ {\rm MK}$, which agrees with the coronal temperature, $T_c \simeq 3 \ {\rm MK}$, associated with X-ray emitting regions \citep{Drake2000} on the stellar surface. Coupled with the detected low circular polarization fraction, this comparison postulates a likely thermal nature of the observed radiation. \\
\begin{figure}[t!]
\centering
\includegraphics[width=0.47\textwidth,, trim={0.25cm 0.1cm 1.2cm 1cm}, clip]{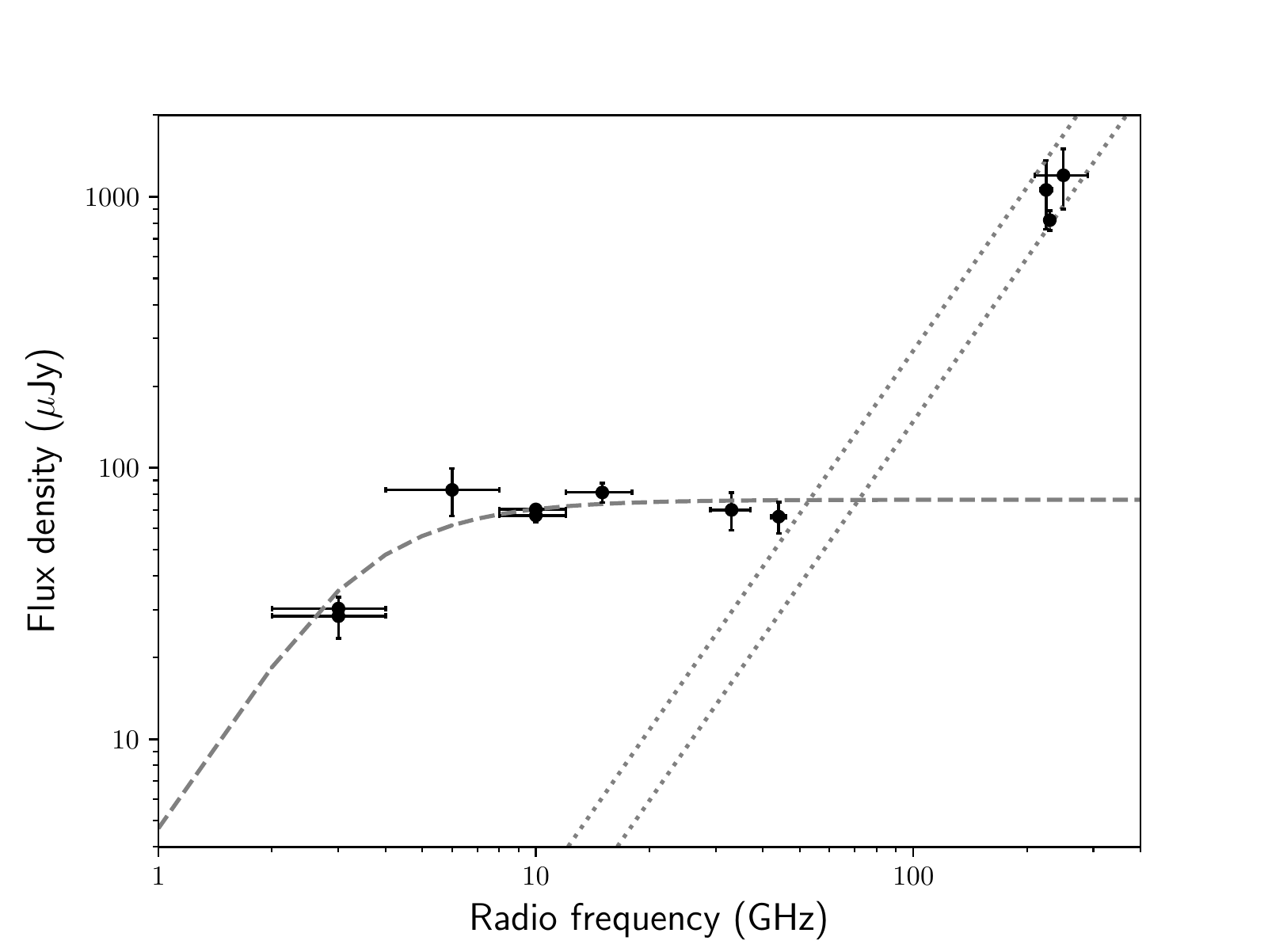}
\caption{2---300~GHz spectrum of continuum emission associated with $\rm \epsilon \ Eri$. The plotted points represent different multi-epoch, multi-frequency flux density estimates listed in Table~\ref{tab1}. The gray dashed line is a smooth spectral fit that is constant above 8~GHz, and follows a $S_{\nu} \propto \nu^2$ scaling below 4---8~GHz. The gray dotted lines indicate $S_{\nu} \propto \nu^2$~blackbody laws for $T \approx 10^4$~K (upper) and $T \approx 5600$~K (lower). Acknowledging possible inter-epoch variability, we refrain from fitting precise power-law models separately over the 2--6, 6---50, and 100---300~GHz frequency ranges. \label{fig4}}
\end{figure}

Figure~\ref{fig4} illustrates the broadband continuum spectrum of $\rm \epsilon \ Eri$ using the multi-frequency flux density estimates listed in Table~\ref{tab1}. Consistent with \citet{Bastian2018}, our findings confirm a sharp decline in continuum flux density below 4---8~GHz. Accounting for likely inter-epoch variability between different spectral measurements, we identify emission processes that explain general spectral trends over different frequency ranges. The millimeter wave spectrum conforms with expectations of an optically thick, thermally emitting stellar chromosphere having $T \approx 8000 \pm 2400$~K \citep{MacGregor2015}. On the other hand, the observed radio spectrum is flat over 6---50~GHz, and follows a $S_{\nu} \propto \nu^2$ scaling below 6~GHz. We note that our in-band spectral measurements at 2---4~GHz (see Figure~\ref{fig3}) agree with our band-averaged flux estimates, and hence, do not argue against a broadband $\nu^2$ scaling below 6~GHz. \citet{Bastian2018} attribute the flat 6---50~GHz spectrum to optically thick, thermal emission from active regions with a frequency-dependent filling factor. On the other hand, \citet{Rodriguez2019} propose that it is optically thin, thermal free-free emission from a stellar wind. Here, in Sections \ref{sec4.1} and \ref{sec4.2}, we discuss the optically thick 2---4~GHz spectrum of $\rm \epsilon \ Eri$ in the context of two plausible sources of thermal bremsstrahlung, respectively, a magnetically confined stellar corona, and an ionized stellar wind. Section \ref{sec4.3} quantifies the significance of our flare non-detection relative to an apparent flaring rate inferred from \citet{Bastian2018}.

\subsection{Coronal and Chromospheric Emissions}\label{sec4.1}

\begin{deluxetable}{lC}
\tablecaption{Summary of photospheric, chromospheric and coronal properties of $\rm \epsilon \ Eri$.\label{tab4}}
\tablewidth{\textwidth}
\tablehead{
\colhead{Property} & \colhead{Value}
}
\startdata
Photospheric escape speed, $v_w$ (km~s$^{-1}$) & 656\\
Chromospheric temperature, $T_{\rm ch}$ (K) & \sim 10^4\\
Coronal temperature, $T_c$ (K) & 3 \times 10^6\\
Coronal filling factor at 2---4~GHz, $f_c(\nu)$ & \approx 40\% \\
Wind electron temperature, $T_e$ (K) & \sim 10^6
\enddata
\tablenotetext{\dagger}{ X-ray luminosity in the $0.2 - 20$ keV band.}
\vspace{-2cm}
\end{deluxetable}

Thermal free-free absorption and thermal gyroresonance absorption constitute the major opacity sources in the stellar corona. The thermal free-free absorption coefficient scales as $\kappa_{\rm ff} \propto n_e^2\nu^{-2}T^{-3/2}$ \citep{Dulk1985}, where $n_e$ is the electron density. Therefore, $\kappa_{\rm ff}(\nu)$ dominates at low frequencies in cool, dense plasma. On the other hand, thermal gyroresonant absorption ($\kappa_{\rm gr} \propto n_e T^{p}B^{q}$, $p, q>$~1) of low harmonics ($\nu=s\nu_{\rm B, e}$,~$s \simeq$~2---5) of the electron cyclotron frequency ($ \nu_{\rm B,e} = 2.8 \ {\rm MHz} \ B_{\rm gauss}$) becomes relevant in hot coronal plasma above active regions, particularly at high frequencies where $\kappa_{\rm ff}(\nu)$ is negligible. At large optical depths ($\tau_{\nu} \gg 1$), $\kappa_{\rm ff}$ and $\kappa_{\rm gr}$ lead to coronal hotspots superposed on a relatively cool stellar disk with chromospheric temperature, $T_{\rm ch} \simeq T_{\rm ch, \odot} \sim 10^4 \ \rm K$ \citep{Jordan1987}. Following \citet{Villadsen2014} and \citet{Bastian2018}, we constrain the filling factor ($f_c$) of such hotspots on the stellar disk using the model:
\begin{align}\label{eqn8}
\TB &= f_c T_c + (1-f_c)T_{\rm ch}.
\end{align}
Taking $T_c \simeq 3$~MK, our observed $\TB \simeq 1.2$~MK necessitates $f_c \approx 40\%$ at 2---4~GHz, which is consistent with the lower $f_c$ estimates obtained by \citet{Bastian2018} at higher frequencies. Above an active region, lower frequencies correspond to weaker magnetic fields higher up from the footpoint. Magnetic flux freezing in the highly conductive coronal plasma then implies a mapping to larger projected areas for weaker fields, resulting in greater $f_c$ at lower frequencies \citep{Lee1998}. Absorption of the $s=3$ harmonic of $ \nu_{\rm B,e}$ then implies $B \simeq 357 \ {\rm G}$ at the coronal heights from where the 2---4~GHz emission emanates.\\

Figure \ref{fig4} shows that the stellar corona of $\rm \epsilon \ Eri$ becomes optically thick below $\sim 6$~GHz. Using constraints from recent X-ray observations \citep{Coffaro2020} of $\epsilon$~Eri, we identify the radio frequency $\nu_{\rm ff}$, below which $\kappa_{\rm ff}(\nu)$ dominates the net opacity. At any radio frequency, the free-free optical depth ($\tau_{\rm ff}$) is given by:
\begin{align}\label{eqn9}
\tau_{\rm ff} (\nu) = & 2 \times 10^{-28} \left( \frac{T}{10^6~{\rm K}} \right)^{-3/2} \left( \frac{\nu}{1~{\rm GHz}}\right)^{-2} \ldots \nonumber \\
& \left( \frac{\rm EMD}{10^{50}~{\rm cm}^{-3}} \right) \left( \frac{R_{*}}{R_{\odot}} \right)^{-2} f_c^{-1}.
\end{align}
Here, ${\rm EMD} = \int n_e^2 \text{d}V$ is the emission measure distribution (EMD) of X-ray emitting coronal volumes. Setting $\rm{EMD} = 3.2 \times 10^{50}$~cm$^{-3}$ at $T\approx 3$~MK \citep{Coffaro2020}, and $f_c \gtrsim 40\%$ below 4~GHz based on magnetic flux freezing, we find that $\tau_{\rm ff}(\nu) = 1$ occurs at $\nu_{\rm ff} \lesssim 1$~GHz. Hence, it is likely that gyroresonance absorption renders the stellar corona optically thick to radiation in the 2---6~GHz frequency band. The observed low circular polarization fraction of the 2---4~GHz continuum emission can then be explained if the stellar corona of $\epsilon$~Eri is optically thick to both the ordinary and extraordinary modes of gryoresonance emission \citep{Vourlidas2006}. In addition, unless active regions on $\epsilon$~Eri preferentially exhibit one hand of magnetic helicity, a low circular polarization is expected for the disk-averaged radio emission.

\subsection{Free-free Emission from an Ionized Stellar Wind}\label{sec4.2}
The diffuse corona tied to open magnetic field lines streams into the interplanetary medium, forming the stellar wind. Radio emission from stars with high mass loss rates \citep{Abbott1981, Leitherer1997, Fichtinger2017} is primarily governed by free-free interactions between charged species in their ionized winds. Analytic forms of standard wind spectra \citep{Olnon1975, Panagia1975, Wright1975} typically presume a spherically symmetric, isothermal, constant velocity ($v_w$) outflow with a steady mass loss rate ($\dot{M}$). These assumptions imply a wind density profile, $n_e (r) \propto r^{-2}$, where $r$ denotes the radial distance from the wind engine. \\

Fitting a standard wind model to the flat $\rm 6-50 \ GHz$ spectrum in Figure \ref{fig4}, \citet{Rodriguez2019} argue that the $\rm 6-50 \ GHz$ continuum radiation from $\rm \epsilon \ Eri$ is consistent with optically thin, free-free emission from a stellar wind with $\dot{M} \simeq 6.6 \times 10^{-11} \ M_{\odot} \ {\rm yr}^{-1}$. This $\dot{M}$ value is about 3300 times that of the Sun, and nearly 110 times larger than corresponding estimates derived by \citet{Wood2002} and \citet{Odert2017} using different methods. While \citet{Wood2002} measured $\dot{M}$ indirectly using atmospheric Ly$\alpha$ absorption lines, \citet{Odert2017} invoked a relation between the flare rate and the mass loss rate from coronal mass ejections to determine $\dot{M}$ for $\epsilon$~Eri. \\

The standard wind spectrum assumed by \citet{Rodriguez2019} predicts $S_{\nu} \propto \nu^{0.6}$ at $\rm \tau_{\nu} \gg 1$. Such scaling is clearly disfavored by our 2---4~GHz data, suggesting minimal wind contribution to the thermal emission from $\rm \epsilon \ Eri$. Allowing for $n_e(r) \propto$~$r^{-\delta}$, our optically thick spectral index ($S_{\nu}\propto \nu^{\alpha}; \ \alpha \simeq2$), requires $\rm \delta \to \infty$ \citep{Panagia1975}. Such steep density distributions are unphysical in stellar winds, where non-spherical geometry resulting from stellar rotation and magnetic fields typically yields flatter density distributions (see \citealt{SchmidBurgk1982} for a discussion on radiative fluxes for non-spherical stellar envelopes). \\

Assuming a well-behaved, thermal free-free emitting wind with solar-like composition, we use our 2---4~GHz data to place a stringent upper limit on $\dot{M}$ for $\rm \epsilon \ Eri$. Equation (24) of \citet{Panagia1975} relates $\dot{M}$ to the optically thick flux density, $S_{\nu}$, via:
\begin{align}\label{eqn10}
\dot{M} = & \ 2.4 \times 10^{-12} \ M_{\odot} \ {\rm yr}^{-1} \left( \frac{S_{\nu}}{1 \ \mu \rm Jy} \right)^{3/4}  \left( \frac{v_w}{v_{\rm esc}} \right) \ldots \nonumber \\
& \left( \frac{d}{3.2 \rm \ pc} \right)^{3/2} \left( \frac{\nu}{3 \ \rm GHz} \right)^{-9/20}  \left( \frac{T_e}{10^6 \ \rm K} \right)^{-3/40}.
\end{align}
As indicated in Table~\ref{tab4}, $v_{\rm esc} \simeq 656 \rm \ km \ s^{-1}$ is the photospheric escape speed of $\rm \epsilon \ Eri$, and $T_e \sim 10^6 \ \rm K$ is the local electron temperature at the base of the wind. Setting the wind velocity, $v_w = v_{\rm esc}$, and $S_{\nu} < 29 \ \mu \rm Jy$ at $\rm \nu = 3 \ GHz$, we obtain $\dot{M} <  3 \times 10^{-11} \ M_{\odot} \ {\rm yr}^{-1}$. The absolute ionized mass loss rate is likely much lower depending on the fractional contribution of wind emission to the observed flux density, and the ion density at different radial distances from $\rm \epsilon \ Eri$. We note that \citet{Cranmer2011} predict a low $\dot{M} \simeq 5 \times 10^{-14} \ M_{\odot}$~yr$^{-1}$ using stellar wind models of cool stars driven by Alfv\'en waves and turbulence. This estimate is only a factor of 2 above the solar value, possibly suggesting that the $\dot{M}$ values derived by \cite{Wood2002} and \citet{Rodriguez2019} may be significant overestimates.

\subsection{Significance of Flare Non-detections}\label{sec4.3}
Stellar flares are often linked to magnetic energy release episodes in the corona, with their event rates highly dependent on stellar activity cycles. Regular monitoring of $\rm \epsilon \ Eri$ over the past 50 years has revealed the simultaneous operation of two dynamos \citep{Metcalfe2013, Coffaro2020} with cycle periods of $\rm 2.95 \ years$ and $\rm 12.7 \ years$. For comparison, the Sun also exhibits dual dynamos with a short $\sim 2$-year-cycle and a long 11-year-cycle \citep{Fletcher2010}. Using 2---4~GHz VLA observations close to the peak of the $\sim$ 3-year cycle of $\rm \epsilon \ Eri$ in 2016, \citet{Bastian2018} reported one $\sim$ mJy flare in a 20-minute scan on $\rm \epsilon \ Eri$. While the exact physical origin of this flare is still uncertain, we use the apparent mean radio flaring rate, $\rm \lambda_{radio} = 0.05 \ min^{-1}$, to predict a flare count for our net 10-hour exposure on $ \rm \epsilon \ Eri$ in 2019. We note that $\lambda_{\rm radio}$ is several orders of magnitude higher than the observed X-ray flaring rate, $\lambda_{\rm X-ray} \simeq 4.6 \times 10^{-7}$~min$^{-1}$, inferred from the detection of four X-ray flares by \citet{Coffaro2020} over 6000~days of observing.\\

Assuming a Poisson distribution of flare numbers with flare rate $\lambda_{\rm radio}$, the expected flare count for our VLA observations is $\rm \left\langle N \right\rangle = 30 \pm 6$. This estimate markedly deviates from our findings in Section \ref{sec3.1} suggesting that the flare detection by \citet{Bastian2018} was highly fortuitous. We note that our above arguments do not incorporate the intrinsic flare flux density distribution and the detection threshold of our data analysis. According to \citet{Nita2002}, the cumulative peak flux density distribution of solar radio bursts at $\nu > 2$~GHz is well described by ${\rm N}(>S_{\nu}) \propto S_{\nu}^{-1.82}$ at both solar maxima and solar minima. However, given the rapid rotation and relatively young age of $\rm \epsilon \ Eri$, a similar flux density distribution may not hold for flares originating on $\rm \epsilon \ Eri$. Moreover, using observations of chromospheric Ca~II levels, \citet{Coffaro2020} recently reported the non-detection of the latest stellar maximum expected in 2019.\\

Chromospheric activity is tightly linked to changes in the stellar magnetic field and consequently, the subsurface convection zone, stellar rotation and magnetic field regeneration via self-sustaining dynamos \citep{Hall2008}. Hence, the non-detection of the most recent peak of the chromospheric Ca~II cycle suggests a likely change in the internal dynamo of $\rm \epsilon \ Eri$. While the coincidence of our flare non-detection with this undetected stellar maximum supports a stellar origin for the \citet{Bastian2018} flare, models of coherent emission produced through star-planet interactions \citep{Turnpenney2018} cannot be discarded.

\section{Summary and Conclusions}\label{sec5}
Using the VLA at 2---4~GHz, we have conducted the deepest radio survey of the $\rm \epsilon \ Eri$ system reported to date. Our observations have revealed a $\rm 29 \ \mu Jy$, compact continuum source coincident with the {\it Gaia} DR2 position of $\rm \epsilon \ Eri$ to within $0\farcs06$ ($\lesssim 2\sigma$). At our sensitivity threshold, we detected neither significant variability, nor circularly polarized emission ($< 25\%$ at $2.5\sigma$ significance) associated with this source. Assuming a source size equal to the stellar radius of $\rm \epsilon \ Eri$, the source $\TB$ is comparable to the coronal temperature on $\rm \epsilon \ Eri$, suggesting a likely thermal nature of the observed emission. \\

Combining our 2---4~GHz data with previous high frequency measurements \citep{MacGregor2015, Bastian2018, Rodriguez2019}, our observations confirm a spectral turnover of the continuum radiation from $\rm \epsilon \ Eri$ in the 4---8~GHz band. We attribute the observed 2---6~GHz emission to thermal gyroresonance radiation from the optically thick stellar corona of $\rm \epsilon \ Eri$. Optically thick, thermal free-free emission from the stellar corona may become relevant at $\nu \lesssim 1$~GHz. \\

The observed steep spectral index ($\alpha \simeq 2$) in the 2---6~GHz band strongly disfavors thermal free-free emission ($S_{\nu} \propto \nu^{0.6}$ at $\tau_{\nu} \gg 1$) from a stellar wind as a plausible interpretation for the 2---50~GHz continuum radiation from $\epsilon$~Eri. Thus, attributing the entire observed flux density at 3~GHz to stellar wind-associated thermal free-free emission, we derive the upper limit, $ \dot{M} < 3 \times 10^{-11} \ M_{\odot} \ {\rm yr}^{-1}$, for $\rm \epsilon \ Eri$ at the time of our observations. \\

Despite scheduling our observations close to the expected maximum of the $\sim$ 3-year magnetic cycle of $\rm \epsilon \ Eri$ in 2019, we detected no flares in our data above our sensitivity limits. A comparison with the mean flaring rate inferred from \citet{Bastian2018} then reveals the extremely fortunate nature of their flare discovery in 2016. However, the optical non-detection \citep{Coffaro2020} of the most recent stellar maximum expected in 2019 hints at a possible evolution of the internal dynamo of $\rm \epsilon \ Eri$. We encourage continued monitoring of $\rm \epsilon \ Eri$ to track its activity levels and better understand its internal dynamo. Periods of high stellar activity should preferably be followed up both at 2---4~GHz and at low radio frequencies ($\rm \nu < 300$~MHz) to discriminate between stellar and planetary emission models for any flare recurrences.

\acknowledgments{
All authors thank J.~Villadsen for initial conversations that motivated this work. AS thanks P.~Nicholson, and R.~J.~Jennings for useful thought-provoking discussions. AS, SC and JMC acknowledge support from the National Science Foundation (AAG 1815242). The VLA observations presented here were obtained as part of program VLA/19A$-$283, PI: A.~Suresh. The VLA is operated by the National Radio Astronomy Observatory, a facility of the National Science Foundation operated under cooperative agreement by Associated Universities, Inc. \\}

\facilities{VLA}
\software{CASA \citep{CASA}, Python 3 (\url{http://www.python.org}), Astropy \citep{astropy}, , NumPy \citep{numpy}, Matplotlib \citep{matplotlib}, SciPy \citep{scipy}} 

\bibliography{references}

\begin{thebibliography}{}
\expandafter\ifx\csname natexlab\endcsname\relax\def\natexlab#1{#1}\fi
\providecommand{\url}[1]{\href{#1}{#1}}
\providecommand{\dodoi}[1]{doi:~\href{http://doi.org/#1}{\nolinkurl{#1}}}
\providecommand{\doeprint}[1]{\href{http://ascl.net/#1}{\nolinkurl{http://ascl.net/#1}}}
\providecommand{\doarXiv}[1]{\href{https://arxiv.org/abs/#1}{\nolinkurl{https://arxiv.org/abs/#1}}}

\bibitem[{{Abbott} {et~al.}(1981){Abbott}, {Bieging}, \&
  {Churchwell}}]{Abbott1981}
{Abbott}, D.~C., {Bieging}, J.~H., \& {Churchwell}, E. 1981, \apj, 250, 645,
  \dodoi{10.1086/159412}

\bibitem[{{Backman} {et~al.}(2009){Backman}, {Marengo}, {Stapelfeldt}, {Su},
  {Wilner}, {Dowell}, {Watson}, {Stansberry}, {Rieke}, {Megeath}, {Fazio}, \&
  {Werner}}]{Backman2009}
{Backman}, D., {Marengo}, M., {Stapelfeldt}, K., {et~al.} 2009, \apj, 690,
  1522, \dodoi{10.1088/0004-637X/690/2/1522}

\bibitem[{{Bastian} {et~al.}(2018){Bastian}, {Villadsen}, {Maps}, {Hallinan},
  \& {Beasley}}]{Bastian2018}
{Bastian}, T.~S., {Villadsen}, J., {Maps}, A., {Hallinan}, G., \& {Beasley},
  A.~J. 2018, \apj, 857, 133, \dodoi{10.3847/1538-4357/aab3cb}

\bibitem[{{Beasley} {et~al.}(2002){Beasley}, {Gordon}, {Peck}, {Petrov},
  {MacMillan}, {Fomalont}, \& {Ma}}]{Beasley2002}
{Beasley}, A.~J., {Gordon}, D., {Peck}, A.~B., {et~al.} 2002, \apjs, 141, 13,
  \dodoi{10.1086/339806}

\bibitem[{{Benedict} {et~al.}(2006){Benedict}, {McArthur}, {Gatewood}, {Nelan},
  {Cochran}, {Hatzes}, {Endl}, {Wittenmyer}, {Baliunas}, {Walker}, {Yang},
  {K{\"u}rster}, {Els}, \& {Paulson}}]{Benedict2006}
{Benedict}, G.~F., {McArthur}, B.~E., {Gatewood}, G., {et~al.} 2006, \aj, 132,
  2206, \dodoi{10.1086/508323}

\bibitem[{{Bonfanti} {et~al.}(2015){Bonfanti}, {Ortolani}, {Piotto}, \&
  {Nascimbeni}}]{Bonfanti2015}
{Bonfanti}, A., {Ortolani}, S., {Piotto}, G., \& {Nascimbeni}, V. 2015, \aap,
  575, A18, \dodoi{10.1051/0004-6361/201424951}

\bibitem[{{Booth} {et~al.}(2017){Booth}, {Dent}, {Jord{\'a}n}, {Lestrade},
  {Hales}, {Wyatt}, {Casassus}, {Ertel}, {Greaves}, {Kennedy}, {Matr{\`a}},
  {Augereau}, \& {Villard}}]{Booth2017}
{Booth}, M., {Dent}, W. R.~F., {Jord{\'a}n}, A., {et~al.} 2017, \mnras, 469,
  3200, \dodoi{10.1093/mnras/stx1072}

\bibitem[{{Briggs}(1995)}]{Briggs1995}
{Briggs}, D.~S. 1995, in American Astronomical Society Meeting Abstracts, Vol.
  187, American Astronomical Society Meeting Abstracts, 112.02

\bibitem[{{Chavez-Dagostino} {et~al.}(2016){Chavez-Dagostino}, {Bertone},
  {Cruz-Saenz de Miera}, {Marshall}, {Wilson}, {S{\'a}nchez-Arg{\"u}elles},
  {Hughes}, {Kennedy}, {Vega}, {De la Luz}, {Dent}, {Eiroa}, {G{\'o}mez-Ruiz},
  {Greaves}, {Lizano}, {L{\'o}pez-Valdivia}, {Mamajek}, {Monta{\~n}a},
  {Olmedo}, {Rodr{\'\i}guez-Montoya}, {Schloerb}, {Yun}, {Zavala}, \&
  {Zeballos}}]{ChavezDagostino2016}
{Chavez-Dagostino}, M., {Bertone}, E., {Cruz-Saenz de Miera}, F., {et~al.}
  2016, \mnras, 462, 2285, \dodoi{10.1093/mnras/stw1363}

\bibitem[{{Chiti} {et~al.}(2016){Chiti}, {Chatterjee}, {Wharton}, {Cordes},
  {Lazio}, {Kaplan}, {Bower}, \& {Croft}}]{Chiti2016}
{Chiti}, A., {Chatterjee}, S., {Wharton}, R., {et~al.} 2016, \apj, 833, 11,
  \dodoi{10.3847/0004-637X/833/1/11}

\bibitem[{{Coffaro} {et~al.}(2020){Coffaro}, {Stelzer}, {Orlando}, {Hall},
  {Metcalfe}, {Wolter}, {Mittag}, {Sanz-Forcada}, {Schneider}, \&
  {Ducci}}]{Coffaro2020}
{Coffaro}, M., {Stelzer}, B., {Orlando}, S., {et~al.} 2020, \aap, 636, A49,
  \dodoi{10.1051/0004-6361/201936479}

\bibitem[{{Cranmer} \& {Saar}(2011)}]{Cranmer2011}
{Cranmer}, S.~R., \& {Saar}, S.~H. 2011, \apj, 741, 54,
  \dodoi{10.1088/0004-637X/741/1/54}

\bibitem[{{Drake}(1961)}]{Drake1961}
{Drake}, F.~D. 1961, Physics Today, 14, 40, \dodoi{10.1063/1.3057500}

\bibitem[{{Drake} {et~al.}(2000){Drake}, {Peres}, {Orlando}, {Laming}, \&
  {Maggio}}]{Drake2000}
{Drake}, J.~J., {Peres}, G., {Orlando}, S., {Laming}, J.~M., \& {Maggio}, A.
  2000, \apj, 545, 1074, \dodoi{10.1086/317820}

\bibitem[{{Dulk}(1985)}]{Dulk1985}
{Dulk}, G.~A. 1985, \araa, 23, 169, \dodoi{10.1146/annurev.aa.23.090185.001125}

\bibitem[{{Fichtinger} {et~al.}(2017){Fichtinger}, {G{\"u}del}, {Mutel},
  {Hallinan}, {Gaidos}, {Skinner}, {Lynch}, \& {Gayley}}]{Fichtinger2017}
{Fichtinger}, B., {G{\"u}del}, M., {Mutel}, R.~L., {et~al.} 2017, \aap, 599,
  A127, \dodoi{10.1051/0004-6361/201629886}

\bibitem[{{Fletcher} {et~al.}(2010){Fletcher}, {Broomhall}, {Salabert}, {Basu},
  {Chaplin}, {Elsworth}, {Garcia}, \& {New}}]{Fletcher2010}
{Fletcher}, S.~T., {Broomhall}, A.-M., {Salabert}, D., {et~al.} 2010, \apjl,
  718, L19, \dodoi{10.1088/2041-8205/718/1/L19}

\bibitem[{{Gaia Collaboration} {et~al.}(2016){Gaia Collaboration}, {Prusti},
  {de Bruijne}, {Brown}, {Vallenari}, {Babusiaux}, {Bailer-Jones}, {Bastian},
  {Biermann}, {Evans}, {Eyer}, {Jansen}, {Jordi}, {Klioner}, {Lammers},
  {Lindegren}, {Luri}, {Mignard}, {Milligan}, {Panem}, {Poinsignon},
  {Pourbaix}, {Randich}, {Sarri}, {Sartoretti}, {Siddiqui}, {Soubiran},
  {Valette}, {van Leeuwen}, {Walton}, {Aerts}, {Arenou}, {Cropper}, {Drimmel},
  {H{\o}g}, {Katz}, {Lattanzi}, {O'Mullane}, {Grebel}, {Holland}, {Huc},
  {Passot}, {Bramante}, {Cacciari}, {Casta{\~n}eda}, {Chaoul}, {Cheek}, {De
  Angeli}, {Fabricius}, {Guerra}, {Hern{\'a}ndez}, {Jean-Antoine-Piccolo},
  {Masana}, {Messineo}, {Mowlavi}, {Nienartowicz}, {Ord{\'o}{\~n}ez-Blanco},
  {Panuzzo}, {Portell}, {Richards}, {Riello}, {Seabroke}, {Tanga},
  {Th{\'e}venin}, {Torra}, {Els}, {Gracia-Abril}, {Comoretto},
  {Garcia-Reinaldos}, {Lock}, {Mercier}, {Altmann}, {Andrae}, {Astraatmadja},
  {Bellas-Velidis}, {Benson}, {Berthier}, {Blomme}, {Busso}, {Carry},
  {Cellino}, {Clementini}, {Cowell}, {Creevey}, {Cuypers}, {Davidson}, {De
  Ridder}, {de Torres}, {Delchambre}, {Dell'Oro}, {Ducourant}, {Fr{\'e}mat},
  {Garc{\'\i}a-Torres}, {Gosset}, {Halbwachs}, {Hambly}, {Harrison}, {Hauser},
  {Hestroffer}, {Hodgkin}, {Huckle}, {Hutton}, {Jasniewicz}, {Jordan},
  {Kontizas}, {Korn}, {Lanzafame}, {Manteiga}, {Moitinho}, {Muinonen},
  {Osinde}, {Pancino}, {Pauwels}, {Petit}, {Recio-Blanco}, {Robin}, {Sarro},
  {Siopis}, {Smith}, {Smith}, {Sozzetti}, {Thuillot}, {van Reeven}, {Viala},
  {Abbas}, {Abreu Aramburu}, {Accart}, {Aguado}, {Allan}, {Allasia},
  {Altavilla}, {{\'A}lvarez}, {Alves}, {Anderson}, {Andrei}, {Anglada Varela},
  {Antiche}, {Antoja}, {Ant{\'o}n}, {Arcay}, {Atzei}, {Ayache}, {Bach},
  {Baker}, {Balaguer-N{\'u}{\~n}ez}, {Barache}, {Barata}, {Barbier}, {Barblan},
  {Baroni}, {Barrado y Navascu{\'e}s}, {Barros}, {Barstow}, {Becciani},
  {Bellazzini}, {Bellei}, {Bello Garc{\'\i}a}, {Belokurov}, {Bendjoya},
  {Berihuete}, {Bianchi}, {Bienaym{\'e}}, {Billebaud}, {Blagorodnova},
  {Blanco-Cuaresma}, {Boch}, {Bombrun}, {Borrachero}, {Bouquillon}, {Bourda},
  {Bouy}, {Bragaglia}, {Breddels}, {Brouillet}, {Br{\"u}semeister},
  {Bucciarelli}, {Budnik}, {Burgess}, {Burgon}, {Burlacu}, {Busonero}, {Buzzi},
  {Caffau}, {Cambras}, {Campbell}, {Cancelliere}, {Cantat-Gaudin}, {Carlucci},
  {Carrasco}, {Castellani}, {Charlot}, {Charnas}, {Charvet}, {Chassat},
  {Chiavassa}, {Clotet}, {Cocozza}, {Collins}, {Collins}, {Costigan}, {Crifo},
  {Cross}, {Crosta}, {Crowley}, {Dafonte}, {Damerdji}, {Dapergolas}, {David},
  {David}, {De Cat}, {de Felice}, {de Laverny}, {De Luise}, {De March}, {de
  Martino}, {de Souza}, {Debosscher}, {del Pozo}, {Delbo}, {Delgado},
  {Delgado}, {di Marco}, {Di Matteo}, {Diakite}, {Distefano}, {Dolding}, {Dos
  Anjos}, {Drazinos}, {Dur{\'a}n}, {Dzigan}, {Ecale}, {Edvardsson}, {Enke},
  {Erdmann}, {Escolar}, {Espina}, {Evans}, {Eynard Bontemps}, {Fabre},
  {Fabrizio}, {Faigler}, {Falc{\~a}o}, {Farr{\`a}s Casas}, {Faye}, {Federici},
  {Fedorets}, {Fern{\'a}ndez-Hern{\'a}ndez}, {Fernique}, {Fienga}, {Figueras},
  {Filippi}, {Findeisen}, {Fonti}, {Fouesneau}, {Fraile}, {Fraser}, {Fuchs},
  {Furnell}, {Gai}, {Galleti}, {Galluccio}, {Garabato}, {Garc{\'\i}a-Sedano},
  {Gar{\'e}}, {Garofalo}, {Garralda}, {Gavras}, {Gerssen}, {Geyer}, {Gilmore},
  {Girona}, {Giuffrida}, {Gomes}, {Gonz{\'a}lez-Marcos},
  {Gonz{\'a}lez-N{\'u}{\~n}ez}, {Gonz{\'a}lez-Vidal}, {Granvik}, {Guerrier},
  {Guillout}, {Guiraud}, {G{\'u}rpide}, {Guti{\'e}rrez-S{\'a}nchez}, {Guy},
  {Haigron}, {Hatzidimitriou}, {Haywood}, {Heiter}, {Helmi}, {Hobbs},
  {Hofmann}, {Holl}, {Holland }, {Hunt}, {Hypki}, {Icardi}, {Irwin}, {Jevardat
  de Fombelle}, {Jofr{\'e}}, {Jonker}, {Jorissen}, {Julbe}, {Karampelas},
  {Kochoska}, {Kohley}, {Kolenberg}, {Kontizas}, {Koposov}, {Kordopatis},
  {Koubsky}, {Kowalczyk}, {Krone-Martins}, {Kudryashova}, {Kull}, {Bachchan},
  {Lacoste-Seris}, {Lanza}, {Lavigne}, {Le Poncin-Lafitte}, {Lebreton},
  {Lebzelter}, {Leccia}, {Leclerc}, {Lecoeur-Taibi}, {Lemaitre}, {Lenhardt},
  {Leroux}, {Liao}, {Licata}, {Lindstr{\o}m}, {Lister}, {Livanou}, {Lobel},
  {L{\"o}ffler}, {L{\'o}pez}, {Lopez-Lozano}, {Lorenz}, {Loureiro},
  {MacDonald}, {Magalh{\~a}es Fernandes}, {Managau}, {Mann}, {Mantelet},
  {Marchal}, {Marchant}, {Marconi}, {Marie}, {Marinoni}, {Marrese},
  {Marschalk{\'o}}, {Marshall}, {Mart{\'\i}n-Fleitas}, {Martino}, {Mary},
  {Matijevi{\v{c}}}, {Mazeh}, {McMillan}, {Messina}, {Mestre}, {Michalik},
  {Millar}, {Miranda}, {Molina}, {Molinaro}, {Molinaro}, {Moln{\'a}r},
  {Moniez}, {Montegriffo}, {Monteiro}, {Mor}, {Mora}, {Morbidelli}, {Morel},
  {Morgenthaler}, {Morley}, {Morris}, {Mulone}, {Muraveva}, {Musella},
  {Narbonne}, {Nelemans}, {Nicastro}, {Noval}, {Ord{\'e}novic},
  {Ordieres-Mer{\'e}}, {Osborne}, {Pagani}, {Pagano}, {Pailler}, {Palacin},
  {Palaversa}, {Parsons}, {Paulsen}, {Pecoraro}, {Pedrosa}, {Pentik{\"a}inen},
  {Pereira}, {Pichon}, {Piersimoni}, {Pineau}, {Plachy}, {Plum}, {Poujoulet},
  {Pr{\v{s}}a}, {Pulone}, {Ragaini}, {Rago}, {Rambaux}, {Ramos-Lerate},
  {Ranalli}, {Rauw}, {Read}, {Regibo}, {Renk}, {Reyl{\'e}}, {Ribeiro},
  {Rimoldini}, {Ripepi}, {Riva}, {Rixon}, {Roelens}, {Romero-G{\'o}mez},
  {Rowell}, {Royer}, {Rudolph}, {Ruiz-Dern}, {Sadowski}, {Sagrist{\`a}
  Sell{\'e}s}, {Sahlmann}, {Salgado}, {Salguero}, {Sarasso}, {Savietto},
  {Schnorhk}, {Schultheis}, {Sciacca}, {Segol}, {Segovia}, {Segransan},
  {Serpell}, {Shih}, {Smareglia}, {Smart}, {Smith}, {Solano}, {Solitro},
  {Sordo}, {Soria Nieto}, {Souchay}, {Spagna}, {Spoto}, {Stampa}, {Steele},
  {Steidelm{\"u}ller}, {Stephenson}, {Stoev}, {Suess}, {S{\"u}veges}, {Surdej},
  {Szabados}, {Szegedi-Elek}, {Tapiador}, {Taris}, {Tauran}, {Taylor},
  {Teixeira}, {Terrett}, {Tingley}, {Trager}, {Turon}, {Ulla}, {Utrilla},
  {Valentini}, {van Elteren}, {Van Hemelryck}, {van Leeuwen}, {Varadi},
  {Vecchiato}, {Veljanoski}, {Via}, {Vicente}, {Vogt}, {Voss}, {Votruba},
  {Voutsinas}, {Walmsley}, {Weiler}, {Weingrill}, {Werner}, {Wevers},
  {Whitehead}, {Wyrzykowski}, {Yoldas}, {{\v{Z}}erjal}, {Zucker}, {Zurbach},
  {Zwitter}, {Alecu}, {Allen}, {Allende Prieto}, {Amorim},
  {Anglada-Escud{\'e}}, {Arsenijevic}, {Azaz}, {Balm}, {Beck}, {Bernstein},
  {Bigot}, {Bijaoui}, {Blasco}, {Bonfigli}, {Bono}, {Boudreault}, {Bressan},
  {Brown}, {Brunet}, {Bunclark}, {Buonanno}, {Butkevich}, {Carret}, {Carrion},
  {Chemin}, {Ch{\'e}reau}, {Corcione}, {Darmigny}, {de Boer}, {de Teodoro}, {de
  Zeeuw}, {Delle Luche}, {Domingues}, {Dubath}, {Fodor}, {Fr{\'e}zouls},
  {Fries}, {Fustes}, {Fyfe}, {Gallardo}, {Gallegos}, {Gardiol}, {Gebran},
  {Gomboc}, {G{\'o}mez}, {Grux}, {Gueguen}, {Heyrovsky}, {Hoar}, {Iannicola},
  {Isasi Parache}, {Janotto}, {Joliet}, {Jonckheere}, {Keil}, {Kim},
  {Klagyivik}, {Klar}, {Knude}, {Kochukhov}, {Kolka}, {Kos}, {Kutka}, {Lainey},
  {LeBouquin}, {Liu}, {Loreggia}, {Makarov}, {Marseille}, {Martayan},
  {Martinez-Rubi}, {Massart}, {Meynadier}, {Mignot}, {Munari}, {Nguyen},
  {Nordlander}, {Ocvirk}, {O'Flaherty}, {Olias Sanz}, {Ortiz}, {Osorio},
  {Oszkiewicz}, {Ouzounis}, {Palmer}, {Park}, {Pasquato}, {Peltzer}, {Peralta},
  {P{\'e}turaud}, {Pieniluoma}, {Pigozzi}, {Poels}, {Prat}, {Prod'homme},
  {Raison}, {Rebordao}, {Risquez}, {Rocca-Volmerange}, {Rosen}, {Ruiz-Fuertes},
  {Russo}, {Sembay}, {Serraller Vizcaino}, {Short}, {Siebert}, {Silva},
  {Sinachopoulos}, {Slezak}, {Soffel}, {Sosnowska}, {Strai{\v{z}}ys}, {ter
  Linden}, {Terrell}, {Theil}, {Tiede}, {Troisi}, {Tsalmantza}, {Tur},
  {Vaccari}, {Vachier}, {Valles}, {Van Hamme}, {Veltz}, {Virtanen}, {Wallut},
  {Wichmann}, {Wilkinson}, {Ziaeepour}, \& {Zschocke}}]{GaiaMission}
{Gaia Collaboration}, {Prusti}, T., {de Bruijne}, J.~H.~J., {et~al.} 2016,
  \aap, 595, A1, \dodoi{10.1051/0004-6361/201629272}

\bibitem[{{Gaia Collaboration} {et~al.}(2018){Gaia Collaboration}, {Brown},
  {Vallenari}, {Prusti}, {de Bruijne}, {Babusiaux}, {Bailer-Jones}, {Biermann},
  {Evans}, {Eyer}, {Jansen}, {Jordi}, {Klioner}, {Lammers}, {Lindegren},
  {Luri}, {Mignard}, {Panem}, {Pourbaix}, {Randich}, {Sartoretti}, {Siddiqui},
  {Soubiran}, {van Leeuwen}, {Walton}, {Arenou}, {Bastian}, {Cropper},
  {Drimmel}, {Katz}, {Lattanzi}, {Bakker}, {Cacciari}, {Casta{\~n}eda},
  {Chaoul}, {Cheek}, {De Angeli}, {Fabricius}, {Guerra}, {Holl}, {Masana},
  {Messineo}, {Mowlavi}, {Nienartowicz}, {Panuzzo}, {Portell}, {Riello},
  {Seabroke}, {Tanga}, {Th{\'e}venin}, {Gracia-Abril}, {Comoretto},
  {Garcia-Reinaldos}, {Teyssier}, {Altmann}, {Andrae}, {Audard},
  {Bellas-Velidis}, {Benson}, {Berthier}, {Blomme}, {Burgess}, {Busso},
  {Carry}, {Cellino}, {Clementini}, {Clotet}, {Creevey}, {Davidson}, {De
  Ridder}, {Delchambre}, {Dell'Oro}, {Ducourant},
  {Fern{\'a}ndez-Hern{\'a}ndez}, {Fouesneau}, {Fr{\'e}mat}, {Galluccio},
  {Garc{\'\i}a-Torres}, {Gonz{\'a}lez-N{\'u}{\~n}ez}, {Gonz{\'a}lez-Vidal},
  {Gosset}, {Guy}, {Halbwachs}, {Hambly}, {Harrison}, {Hern{\'a}ndez},
  {Hestroffer}, {Hodgkin}, {Hutton}, {Jasniewicz}, {Jean-Antoine-Piccolo},
  {Jordan}, {Korn}, {Krone-Martins}, {Lanzafame}, {Lebzelter}, {L{\"o}ffler},
  {Manteiga}, {Marrese}, {Mart{\'\i}n-Fleitas}, {Moitinho}, {Mora}, {Muinonen},
  {Osinde}, {Pancino}, {Pauwels}, {Petit}, {Recio-Blanco}, {Richards},
  {Rimoldini}, {Robin}, {Sarro}, {Siopis}, {Smith}, {Sozzetti}, {S{\"u}veges},
  {Torra}, {van Reeven}, {Abbas}, {Abreu Aramburu}, {Accart}, {Aerts},
  {Altavilla}, {{\'A}lvarez}, {Alvarez}, {Alves}, {Anderson}, {Andrei},
  {Anglada Varela}, {Antiche}, {Antoja}, {Arcay}, {Astraatmadja}, {Bach},
  {Baker}, {Balaguer-N{\'u}{\~n}ez}, {Balm}, {Barache}, {Barata}, {Barbato},
  {Barblan}, {Barklem}, {Barrado}, {Barros}, {Barstow}, {Bartholom{\'e}
  Mu{\~n}oz}, {Bassilana}, {Becciani}, {Bellazzini}, {Berihuete}, {Bertone},
  {Bianchi}, {Bienaym{\'e}}, {Blanco-Cuaresma}, {Boch}, {Boeche}, {Bombrun},
  {Borrachero}, {Bossini}, {Bouquillon}, {Bourda}, {Bragaglia}, {Bramante},
  {Breddels}, {Bressan}, {Brouillet}, {Br{\"u}semeister}, {Brugaletta},
  {Bucciarelli}, {Burlacu}, {Busonero}, {Butkevich}, {Buzzi}, {Caffau},
  {Cancelliere}, {Cannizzaro}, {Cantat-Gaudin}, {Carballo}, {Carlucci},
  {Carrasco}, {Casamiquela}, {Castellani}, {Castro-Ginard}, {Charlot},
  {Chemin}, {Chiavassa}, {Cocozza}, {Costigan}, {Cowell}, {Crifo}, {Crosta},
  {Crowley}, {Cuypers}, {Dafonte}, {Damerdji}, {Dapergolas}, {David}, {David},
  {de Laverny}, {De Luise}, {De March}, {de Martino}, {de Souza}, {de Torres},
  {Debosscher}, {del Pozo}, {Delbo}, {Delgado}, {Delgado}, {Di Matteo},
  {Diakite}, {Diener}, {Distefano}, {Dolding}, {Drazinos}, {Dur{\'a}n},
  {Edvardsson}, {Enke}, {Eriksson}, {Esquej}, {Eynard Bontemps}, {Fabre},
  {Fabrizio}, {Faigler}, {Falc{\~a}o}, {Farr{\`a}s Casas}, {Federici},
  {Fedorets}, {Fernique}, {Figueras}, {Filippi}, {Findeisen}, {Fonti},
  {Fraile}, {Fraser}, {Fr{\'e}zouls}, {Gai}, {Galleti}, {Garabato},
  {Garc{\'\i}a-Sedano}, {Garofalo}, {Garralda}, {Gavel}, {Gavras}, {Gerssen},
  {Geyer}, {Giacobbe}, {Gilmore}, {Girona}, {Giuffrida}, {Glass}, {Gomes},
  {Granvik}, {Gueguen}, {Guerrier}, {Guiraud}, {Guti{\'e}rrez-S{\'a}nchez},
  {Haigron}, {Hatzidimitriou}, {Hauser}, {Haywood}, {Heiter}, {Helmi}, {Heu},
  {Hilger}, {Hobbs}, {Hofmann}, {Holland}, {Huckle}, {Hypki}, {Icardi},
  {Jan{\ss}en}, {Jevardat de Fombelle}, {Jonker}, {Juh{\'a}sz}, {Julbe},
  {Karampelas}, {Kewley}, {Klar}, {Kochoska}, {Kohley}, {Kolenberg},
  {Kontizas}, {Kontizas}, {Koposov}, {Kordopatis}, {Kostrzewa-Rutkowska},
  {Koubsky}, {Lambert}, {Lanza}, {Lasne}, {Lavigne}, {Le Fustec}, {Le
  Poncin-Lafitte}, {Lebreton}, {Leccia}, {Leclerc}, {Lecoeur-Taibi},
  {Lenhardt}, {Leroux}, {Liao}, {Licata}, {Lindstr{\o}m}, {Lister}, {Livanou},
  {Lobel}, {L{\'o}pez}, {Managau}, {Mann}, {Mantelet}, {Marchal}, {Marchant},
  {Marconi}, {Marinoni}, {Marschalk{\'o}}, {Marshall}, {Martino}, {Marton},
  {Mary}, {Massari}, {Matijevi{\v{c}}}, {Mazeh}, {McMillan}, {Messina},
  {Michalik}, {Millar}, {Molina}, {Molinaro}, {Moln{\'a}r}, {Montegriffo},
  {Mor}, {Morbidelli}, {Morel}, {Morris}, {Mulone}, {Muraveva}, {Musella},
  {Nelemans}, {Nicastro}, {Noval}, {O'Mullane}, {Ord{\'e}novic},
  {Ord{\'o}{\~n}ez-Blanco}, {Osborne}, {Pagani}, {Pagano}, {Pailler},
  {Palacin}, {Palaversa}, {Panahi}, {Pawlak}, {Piersimoni}, {Pineau}, {Plachy},
  {Plum}, {Poggio}, {Poujoulet}, {Pr{\v{s}}a}, {Pulone}, {Racero}, {Ragaini},
  {Rambaux}, {Ramos-Lerate}, {Regibo}, {Reyl{\'e}}, {Riclet}, {Ripepi}, {Riva},
  {Rivard}, {Rixon}, {Roegiers}, {Roelens}, {Romero-G{\'o}mez}, {Rowell},
  {Royer}, {Ruiz-Dern}, {Sadowski}, {Sagrist{\`a} Sell{\'e}s}, {Sahlmann},
  {Salgado}, {Salguero}, {Sanna}, {Santana-Ros}, {Sarasso}, {Savietto},
  {Schultheis}, {Sciacca}, {Segol}, {Segovia}, {S{\'e}gransan}, {Shih},
  {Siltala}, {Silva}, {Smart}, {Smith}, {Solano}, {Solitro}, {Sordo}, {Soria
  Nieto}, {Souchay}, {Spagna}, {Spoto}, {Stampa}, {Steele},
  {Steidelm{\"u}ller}, {Stephenson}, {Stoev}, {Suess}, {Surdej}, {Szabados},
  {Szegedi-Elek}, {Tapiador}, {Taris}, {Tauran}, {Taylor}, {Teixeira},
  {Terrett}, {Teyssand ier}, {Thuillot}, {Titarenko}, {Torra Clotet}, {Turon},
  {Ulla}, {Utrilla}, {Uzzi}, {Vaillant}, {Valentini}, {Valette}, {van Elteren},
  {Van Hemelryck}, {van Leeuwen}, {Vaschetto}, {Vecchiato}, {Veljanoski},
  {Viala}, {Vicente}, {Vogt}, {von Essen}, {Voss}, {Votruba}, {Voutsinas},
  {Walmsley}, {Weiler}, {Wertz}, {Wevers}, {Wyrzykowski}, {Yoldas},
  {{\v{Z}}erjal}, {Ziaeepour}, {Zorec}, {Zschocke}, {Zucker}, {Zurbach}, \&
  {Zwitter}}]{GaiaDR2}
{Gaia Collaboration}, {Brown}, A.~G.~A., {Vallenari}, A., {et~al.} 2018, \aap,
  616, A1, \dodoi{10.1051/0004-6361/201833051}

\bibitem[{{Gary} \& {Linsky}(1981)}]{Gary1981}
{Gary}, D.~E., \& {Linsky}, J.~L. 1981, \apj, 250, 284, \dodoi{10.1086/159373}

\bibitem[{{Greaves} {et~al.}(2014){Greaves}, {Sibthorpe}, {Acke}, {Pantin},
  {Vandenbussche}, {Olofsson}, {Dominik}, {Barlow}, {Bendo}, {Blommaert},
  {Brand eker}, {de Vries}, {Dent}, {Di Francesco}, {Fridlund}, {Gear},
  {Harvey}, {Hogerheijde}, {Holland}, {Ivison}, {Liseau}, {Matthews},
  {Pilbratt}, {Walker}, \& {Waelkens}}]{Greaves2014}
{Greaves}, J.~S., {Sibthorpe}, B., {Acke}, B., {et~al.} 2014, \apjl, 791, L11,
  \dodoi{10.1088/2041-8205/791/1/L11}

\bibitem[{{Hall}(2008)}]{Hall2008}
{Hall}, J.~C. 2008, Living Reviews in Solar Physics, 5, 2,
  \dodoi{10.12942/lrsp-2008-2}

\bibitem[{{Hatzes} {et~al.}(2000){Hatzes}, {Cochran}, {McArthur}, {Baliunas},
  {Walker}, {Campbell}, {Irwin}, {Yang}, {K{\"u}rster}, {Endl}, {Els},
  {Butler}, \& {Marcy}}]{Hatzes2000}
{Hatzes}, A.~P., {Cochran}, W.~D., {McArthur}, B., {et~al.} 2000, \apjl, 544,
  L145, \dodoi{10.1086/317319}

\bibitem[{{Henry} {et~al.}(1995){Henry}, {Soderblom}, {Baliunas}, {Davis},
  {Donahue}, {Latham}, {Stefanik}, {Torres}, {Duquennoy}, {Mayor}, {Andersen},
  {Nordstrom}, \& {Olsen}}]{Henry1995}
{Henry}, T., {Soderblom}, D., {Baliunas}, S., {et~al.} 1995, Astronomical
  Society of the Pacific Conference Series, Vol.~74, {The Current State of
  Target Selection for NASA's High Resolution Microwave Survey}, ed. G.~S.
  {Shostak}, 207

\bibitem[{{Hunter}(2007)}]{matplotlib}
{Hunter}, J.~D. 2007, Computing in Science Engineering, 9, 90,
  \dodoi{10.1109/MCSE.2007.55}

\bibitem[{{Johnson}(1981)}]{Johnson1981}
{Johnson}, H.~M. 1981, \apj, 243, 234, \dodoi{10.1086/158589}

\bibitem[{{Jordan} {et~al.}(1987){Jordan}, {Ayres}, {Brown}, {Linsky}, \&
  {Simon}}]{Jordan1987}
{Jordan}, C., {Ayres}, T.~R., {Brown}, A., {Linsky}, J.~L., \& {Simon}, T.
  1987, \mnras, 225, 903, \dodoi{10.1093/mnras/225.4.903}

\bibitem[{{Keenan} \& {McNeil}(1989)}]{Keenan1989}
{Keenan}, P.~C., \& {McNeil}, R.~C. 1989, \apjs, 71, 245,
  \dodoi{10.1086/191373}

\bibitem[{{Lee} {et~al.}(1998){Lee}, {McClymont}, {Miki{\'c}}, {White}, \&
  {Kundu}}]{Lee1998}
{Lee}, J., {McClymont}, A.~N., {Miki{\'c}}, Z., {White}, S.~M., \& {Kundu},
  M.~R. 1998, \apj, 501, 853, \dodoi{10.1086/305851}

\bibitem[{{Lehmann} {et~al.}(2015){Lehmann}, {K{\"u}nstler}, {Carroll}, \&
  {Strassmeier}}]{Lehmann2015}
{Lehmann}, L.~T., {K{\"u}nstler}, A., {Carroll}, T.~A., \& {Strassmeier}, K.~G.
  2015, Astronomische Nachrichten, 336, 258, \dodoi{10.1002/asna.201412162}

\bibitem[{{Leitherer} {et~al.}(1997){Leitherer}, {Chapman}, \&
  {Koribalski}}]{Leitherer1997}
{Leitherer}, C., {Chapman}, J.~M., \& {Koribalski}, B. 1997, \apj, 481, 898,
  \dodoi{10.1086/304096}

\bibitem[{{Lestrade} \& {Thilliez}(2015)}]{Lestrade2015}
{Lestrade}, J.-F., \& {Thilliez}, E. 2015, \aap, 576, A72,
  \dodoi{10.1051/0004-6361/201425422}

\bibitem[{{MacGregor} {et~al.}(2015){MacGregor}, {Wilner}, {Andrews},
  {Lestrade}, \& {Maddison}}]{MacGregor2015}
{MacGregor}, M.~A., {Wilner}, D.~J., {Andrews}, S.~M., {Lestrade}, J.-F., \&
  {Maddison}, S. 2015, \apj, 809, 47, \dodoi{10.1088/0004-637X/809/1/47}

\bibitem[{{Mamajek} \& {Hillenbrand}(2008)}]{Mamajek2008}
{Mamajek}, E.~E., \& {Hillenbrand}, L.~A. 2008, \apj, 687, 1264,
  \dodoi{10.1086/591785}

\bibitem[{{Mawet} {et~al.}(2019){Mawet}, {Hirsch}, {Lee}, {Ruffio}, {Bottom},
  {Fulton}, {Absil}, {Beichman}, {Bowler}, {Bryan}, {Choquet}, {Ciardi},
  {Christiaens}, {Defr{\`e}re}, {Gomez Gonzalez}, {Howard}, {Huby}, {Isaacson},
  {Jensen-Clem}, {Kosiarek}, {Marcy}, {Meshkat}, {Petigura}, {Reggiani},
  {Ruane}, {Serabyn}, {Sinukoff}, {Wang}, {Weiss}, \& {Ygouf}}]{Mawet2019}
{Mawet}, D., {Hirsch}, L., {Lee}, E.~J., {et~al.} 2019, \aj, 157, 33,
  \dodoi{10.3847/1538-3881/aaef8a}

\bibitem[{{McMullin} {et~al.}(2007){McMullin}, {Waters}, {Schiebel}, {Young},
  \& {Golap}}]{CASA}
{McMullin}, J.~P., {Waters}, B., {Schiebel}, D., {Young}, W., \& {Golap}, K.
  2007, Astronomical Society of the Pacific Conference Series, Vol. 376, {CASA
  Architecture and Applications}, ed. R.~A. {Shaw}, F.~{Hill}, \& D.~J. {Bell},
  127

\bibitem[{{Metcalfe} {et~al.}(2013){Metcalfe}, {Buccino}, {Brown}, {Mathur},
  {Soderblom}, {Henry}, {Mauas}, {Petrucci}, {Hall}, \& {Basu}}]{Metcalfe2013}
{Metcalfe}, T.~S., {Buccino}, A.~P., {Brown}, B.~P., {et~al.} 2013, \apjl, 763,
  L26, \dodoi{10.1088/2041-8205/763/2/L26}

\bibitem[{{Nita} {et~al.}(2002){Nita}, {Gary}, {Lanzerotti}, \&
  {Thomson}}]{Nita2002}
{Nita}, G.~M., {Gary}, D.~E., {Lanzerotti}, L.~J., \& {Thomson}, D.~J. 2002,
  \apj, 570, 423, \dodoi{10.1086/339577}

\bibitem[{{Odert} {et~al.}(2017){Odert}, {Leitzinger}, {Hanslmeier}, \&
  {Lammer}}]{Odert2017}
{Odert}, P., {Leitzinger}, M., {Hanslmeier}, A., \& {Lammer}, H. 2017, \mnras,
  472, 876, \dodoi{10.1093/mnras/stx1969}

\bibitem[{{Olnon}(1975)}]{Olnon1975}
{Olnon}, F.~M. 1975, \aap, 39, 217

\bibitem[{{Panagia} \& {Felli}(1975)}]{Panagia1975}
{Panagia}, N., \& {Felli}, M. 1975, \aap, 39, 1

\bibitem[{{Perley} \& {Butler}(2017)}]{Perley2017}
{Perley}, R.~A., \& {Butler}, B.~J. 2017, \apjs, 230, 7,
  \dodoi{10.3847/1538-4365/aa6df9}

\bibitem[{{Price-Whelan} {et~al.}(2018){Price-Whelan}, {Sip{\H{o}}cz},
  {G{\"u}nther}, {Lim}, {Crawford}, {Conseil}, {Shupe}, {Craig}, {Dencheva},
  {Ginsburg}, {VanderPlas}, {Bradley}, {P{\'e}rez-Su{\'a}rez}, {de Val-Borro},
  {Paper Contributors}, {Aldcroft}, {Cruz}, {Robitaille}, {Tollerud},
  {Coordination Committee}, {Ardelean}, {Babej}, {Bach}, {Bachetti}, {Bakanov},
  {Bamford}, {Barentsen}, {Barmby}, {Baumbach}, {Berry}, {Biscani}, {Boquien},
  {Bostroem}, {Bouma}, {Brammer}, {Bray}, {Breytenbach}, {Buddelmeijer},
  {Burke}, {Calderone}, {Cano Rodr{\'\i}guez}, {Cara}, {Cardoso}, {Cheedella},
  {Copin}, {Corrales}, {Crichton}, {D{\textquoteright}Avella}, {Deil},
  {Depagne}, {Dietrich}, {Donath}, {Droettboom}, {Earl}, {Erben}, {Fabbro},
  {Ferreira}, {Finethy}, {Fox}, {Garrison}, {Gibbons}, {Goldstein}, {Gommers},
  {Greco}, {Greenfield}, {Groener}, {Grollier}, {Hagen}, {Hirst}, {Homeier},
  {Horton}, {Hosseinzadeh}, {Hu}, {Hunkeler}, {Ivezi{\'c}}, {Jain}, {Jenness},
  {Kanarek}, {Kendrew}, {Kern}, {Kerzendorf}, {Khvalko}, {King}, {Kirkby},
  {Kulkarni}, {Kumar}, {Lee}, {Lenz}, {Littlefair}, {Ma}, {Macleod},
  {Mastropietro}, {McCully}, {Montagnac}, {Morris}, {Mueller}, {Mumford},
  {Muna}, {Murphy}, {Nelson}, {Nguyen}, {Ninan}, {N{\"o}the}, {Ogaz}, {Oh},
  {Parejko}, {Parley}, {Pascual}, {Patil}, {Patil}, {Plunkett}, {Prochaska},
  {Rastogi}, {Reddy Janga}, {Sabater}, {Sakurikar}, {Seifert}, {Sherbert},
  {Sherwood-Taylor}, {Shih}, {Sick}, {Silbiger}, {Singanamalla}, {Singer},
  {Sladen}, {Sooley}, {Sornarajah}, {Streicher}, {Teuben}, {Thomas},
  {Tremblay}, {Turner}, {Terr{\'o}n}, {van Kerkwijk}, {de la Vega}, {Watkins},
  {Weaver}, {Whitmore}, {Woillez}, {Zabalza}, \& {Contributors}}]{astropy}
{Price-Whelan}, A.~M., {Sip{\H{o}}cz}, B.~M., {G{\"u}nther}, H.~M., {et~al.}
  2018, \aj, 156, 123, \dodoi{10.3847/1538-3881/aabc4f}

\bibitem[{{Quillen} \& {Thorndike}(2002)}]{Quillen2002}
{Quillen}, A.~C., \& {Thorndike}, S. 2002, \apjl, 578, L149,
  \dodoi{10.1086/344708}

\bibitem[{{Rodr{\'\i}guez} {et~al.}(2019){Rodr{\'\i}guez}, {Lizano}, {Loinard},
  {Ch{\'a}vez-Dagostino}, {Bastian}, \& {Beasley}}]{Rodriguez2019}
{Rodr{\'\i}guez}, L.~F., {Lizano}, S., {Loinard}, L., {et~al.} 2019, \apj, 871,
  172, \dodoi{10.3847/1538-4357/aaf9a6}

\bibitem[{{Schmid-Burgk}(1982)}]{SchmidBurgk1982}
{Schmid-Burgk}, J. 1982, \aap, 108, 169

\bibitem[{{Su} {et~al.}(2017){Su}, {De Buizer}, {Rieke}, {Krivov}, {L{\"o}hne},
  {Marengo}, {Stapelfeldt}, {Ballering}, \& {Vacca}}]{Su2017}
{Su}, K. Y.~L., {De Buizer}, J.~M., {Rieke}, G.~H., {et~al.} 2017, \aj, 153,
  226, \dodoi{10.3847/1538-3881/aa696b}

\bibitem[{{Tarter}(1996)}]{Tarter1996}
{Tarter}, J.~C. 1996, Society of Photo-Optical Instrumentation Engineers (SPIE)
  Conference Series, Vol. 2704, {Project Phoenix: the Australian deployment},
  ed. S.~A. {Kingsley} \& G.~A. {Lemarchand}, 24--34, \dodoi{10.1117/12.243444}

\bibitem[{{Turnpenney} {et~al.}(2018){Turnpenney}, {Nichols}, {Wynn}, \&
  {Burleigh}}]{Turnpenney2018}
{Turnpenney}, S., {Nichols}, J.~D., {Wynn}, G.~A., \& {Burleigh}, M.~R. 2018,
  \apj, 854, 72, \dodoi{10.3847/1538-4357/aaa59c}

\bibitem[{{van der Walt} {et~al.}(2011){van der Walt}, {Colbert}, \&
  {Varoquaux}}]{numpy}
{van der Walt}, S., {Colbert}, S.~C., \& {Varoquaux}, G. 2011, Computing in
  Science Engineering, 13, 22, \dodoi{10.1109/MCSE.2011.37}

\bibitem[{{Villadsen} {et~al.}(2014){Villadsen}, {Hallinan}, {Bourke},
  {G{\"u}del}, \& {Rupen}}]{Villadsen2014}
{Villadsen}, J., {Hallinan}, G., {Bourke}, S., {G{\"u}del}, M., \& {Rupen}, M.
  2014, \apj, 788, 112, \dodoi{10.1088/0004-637X/788/2/112}

\bibitem[{{Virtanen} {et~al.}(2019){Virtanen}, {Gommers}, {Oliphant},
  {Haberland}, {Reddy}, {Cournapeau}, {Burovski}, {Peterson}, {Weckesser},
  {Bright}, {van der Walt}, {Brett}, {Wilson}, {Jarrod Millman}, {Mayorov},
  {Nelson}, {Jones}, {Kern}, {Larson}, {Carey}, {Polat}, {Feng}, {Moore}, {Vand
  erPlas}, {Laxalde}, {Perktold}, {Cimrman}, {Henriksen}, {Quintero}, {Harris},
  {Archibald}, {Ribeiro}, {Pedregosa}, {van Mulbregt}, \&
  {Contributors}}]{scipy}
{Virtanen}, P., {Gommers}, R., {Oliphant}, T.~E., {et~al.} 2019, arXiv
  e-prints, arXiv:1907.10121.
\newblock \doarXiv{1907.10121}

\bibitem[{{Vourlidas} {et~al.}(2006){Vourlidas}, {Gary}, \&
  {Shibasaki}}]{Vourlidas2006}
{Vourlidas}, A., {Gary}, D.~E., \& {Shibasaki}, K. 2006, \pasj, 58, 11,
  \dodoi{10.1093/pasj/58.1.11}

\bibitem[{{Wood} {et~al.}(2002){Wood}, {M{\"u}ller}, {Zank}, \&
  {Linsky}}]{Wood2002}
{Wood}, B.~E., {M{\"u}ller}, H.-R., {Zank}, G.~P., \& {Linsky}, J.~L. 2002,
  \apj, 574, 412, \dodoi{10.1086/340797}

\bibitem[{{Wright} \& {Barlow}(1975)}]{Wright1975}
{Wright}, A.~E., \& {Barlow}, M.~J. 1975, \mnras, 170, 41,
  \dodoi{10.1093/mnras/170.1.41}

\end{thebibliography}
\bibliographystyle{aasjournal}



\end{document}